\documentclass[11pt,a4paper]{article}
\pdfoutput=1

\usepackage{epsfig}
\usepackage{latexsym}
\usepackage{amsfonts}
\usepackage{amsmath}
\usepackage{amsthm}
\usepackage{amssymb}
\usepackage{amsbsy}
\usepackage{multirow}
\usepackage{slashed}
\usepackage{color}
\usepackage{mathrsfs}
\usepackage{subfigure}
\usepackage[font={footnotesize,sl},bf]{caption}
\usepackage{wrapfig}

\usepackage{jheppub}

\newcommand{\bQ}{Q}
\newcommand{\pQ}{q}
\newcommand{\bN}{N}
\newcommand{\pN}{N^{\rm tube}}
\newcommand{\pn}{\eta}

\def\calb         {{\cal B}}

\def\cale         {{\cal E}}

\def\calh         {{\cal H}}

\def\call         {{\cal L}}

\newsavebox{\uuunit}
\sbox{\uuunit}
    {\setlength{\unitlength}{0.825em}
     \begin{picture}(0.6,0.7)
        \thinlines
        \put(0,0){\line(1,0){0.5}}
        \put(0.15,0){\line(0,1){0.7}}
        \put(0.35,0){\line(0,1){0.8}}
       \multiput(0.3,0.8)(-0.04,-0.02){12}{\rule{0.5pt}{0.5pt}}
     \end {picture}}

\def\be{\begin{equation}}
\def\ee{\end{equation}}
\def\bea{\begin{eqnarray}}
\def\eea{\end{eqnarray}}

\newcommand{\beq}{\begin{eqnarray}}
\newcommand{\eeq}{\end{eqnarray}}

\newcommand{\eal}[1]{\be \begin{aligned} #1 \end{aligned}\ee}

\def\a{\alpha}

\def\d{\delta}

\def\f{\phi}

\newcommand{\ti}[1]{\tilde #1}

\def\pa{\partial}

\def\to{\rightarrow}

\def\sF{{{ F}\!\!\!\!\hskip.8pt\hbox{\raise1pt\hbox{/}}\,}}
\def\som{{{ \omega}\!\!\!\!\hskip.8pt\hbox{\raise1pt\hbox{/}}\,}}
\def\sJ{{{\rm J}\!\!\!\!\hskip.8pt\hbox{\raise1pt\hbox{/}}\,}}

\newcommand{\bdm}{\begin{displaymath}}
\newcommand{\edm}{\end{displaymath}}
\newcommand{\nn}{\nonumber \\}
\renewcommand{\f}[2]{\frac{#1}{#2}}
\newcommand{\bref}[1]{(\ref{#1})}
\newcommand\h{\frac{1}{2}}
\newcommand{\hrho}{\hat{\rho}}

\title{New instability of non-extremal black holes:\\ spitting out supertubes}
\author[a]{Borun D.\ Chowdhury,}
\author[b]{Bert Vercnocke}
\affiliation[a]{Institute for Theoretical Physics, University of Amsterdam,\\
Science Park 904, Postbus 94485, 1090 GL Amsterdam, The Netherlands}
\affiliation[b]{Institut de Physique Th\'eorique, CEA Saclay,\\  91191 Gif sur Yvette, France}

\emailAdd{b.d.chowdhury@uva.nl}
\emailAdd{bert.vercnocke@cea.fr}
\abstract{We search for  stable bound states of non-extremal rotating three-charge black holes in five dimensions (Cvetic-Youm black holes) and supertubes. We do this by studying the potential of supertube probes in the non-extremal black hole background and find that generically the marginally bound state of the supersymmetric limit becomes metastable and disappears with non-extremality (higher temperature). However near extremality there is a range of parameters allowing for stable bound states, which have lower energy than the supertube-black hole merger. Angular momentum is crucial for this effect. We use this setup in the D1-D5 decoupling limit to map a thermodynamic instability of the CFT (a new phase which is entropically dominant over the black hole phase) to a tunneling instability of the black hole towards the supertube-black hole bound state. This generalizes the results of \cite{Bena:2011zw}, which mapped an entropy enigma in the bulk to the dual CFT in a supersymmetric setup.}

\arxivnumber{1010.5641}

\preprint{ITFA 11-13\\IPhT-T11/203}

\begin{document}
\maketitle

\section{Introduction}

In gravitational physics there are  examples where thermodynamic instabilities lead to dynamical instabilities. These include  the Penrose process, superradiance \cite{zel2,Misner:1972kx,staro1,Bekenstein:1973mi,Press:1973zz,Unruh:1974bw} and the Gregory Laflamme instability \cite{Gregory:1993vy,Gregory:1994bj}. In the Penrose process, a particle falling into a rotating black hole splits just outside the horizon with one part  falling into the horizon and the other going off to infinity. The outgoing part has more energy and angular momentum than the infalling particle, reducing the angular momentum of the black hole. Superradiance is the wave analogue of the Penrose process where a wave scattering off a rotating black hole is reflected with a larger amplitude. These processes signal a thermodynamical instability, in the sense that the entropy of the black hole increases in the process. In addition, it was shown in \cite{Dias:2007nj,Chowdhury:2010ct} that, for certain three charge black holes in string theory, during superradiance there is an increase in the dual field theory entropy. Gregory and Laflamme showed that in the presence of a compact direction, on the one hand, a black string wrapping the compact direction is favored entropically for small radius while a black hole is favored for large radius and that on the other hand black strings develop a dynamical instability when the radius is above a threshold radius (`Gregory-Laflamme instability'). In \cite{Chowdhury:2006qn} it was shown that  when a dual field theory can be defined, the Gregory-Laflamme instability  maps to a thermodynamic instability in the field theory. These examples suggest that thermodynamic instabilities lead to dynamics in gravitational physics.\footnote{A recent proposal by Verlinde \cite{Verlinde:2010hp} goes one step further and states that all gravitational force is in fact of an entropic nature.}  

The purpose of this paper is to study a similar phenomenon in the background of the five-dimensional non-extremal rotating black hole (Cvetic-Youm black hole \cite{Cvetic:1996xz}), by studying the potential of probe supertubes in the Cvetic-Youm background. We show that, near extremality, the supertube can form a bound state with the black hole. This state can be either metastable or stable (higher or lower energy than the state where the supertube sits at the black hole horizon).

\medskip

The first motivation of this paper is to understand the nature of the instability of \cite{Bena:2011zw} in the non-extremal context. In that article, a supersymmetric thermodynamic instability of  field theory was mapped to a supersymmetric thermodynamic instability in the dual bulk.  The instability in the bulk corresponds to a bound state of a supertube and the supersymmetric five-dimensional rotating three charge black hole (BMPV \cite{Breckenridge:1996is}) being entropically favored over a single BMPV black hole when the angular momentum of the black hole is large. However, since the supersymmetric black hole and the supersymmetric bound state have the same energy, the thermodynamic instability does {\em not} imply that the highly rotating black hole would emit a supertube: there is no dynamical instability.

The situation for non-extremal rotating black holes promises to be more interesting. As different configurations can have different energies in addition to having different entropies, a thermodynamic instability of the field theory should map to a dynamical instability on the bulk side\footnote{In this article the instabilities we find always involve a supertube tunneling through a barrier and thus the dynamic instabilities are quantum mechanical}. Motivated by the results of \cite{Bena:2011zw}  we conjecture that there is such a process of Penrose/superradiance type for the emission of supertubes from rotating non-extremal black holes. Verifying this claim involves studying non-extremal multi-center solutions. Even though many powerful techniques exist for the construction of supersymmetric multi-centered black hole solutions \cite{Denef:2000nb,Bates:2003vx,Bena:2004de, Gutowski:2004yv,Gauntlett:2004qy,Elvang:2004ds,Bena:2005ni, Bena:2005va,
Berglund:2005vb,Saxena:2005uk},  and recent remarkable progress has also been made in construction techniques for multi-centered non-supersymmetric but still extremal solutions \cite{Goldstein:2008fq,Galli:2009bj,Bena:2009en,Bena:2009ev,Bena:2009fi,Bossard:2009my,Bossard:2009mz,Dall'Agata:2010dy,Bossard:2011kz}, there are no generic fully backreacted multi-centered non-extremal solutions known, so we have to rely on the probe limit to study them.\footnote{The only non-extremal multi-center fuzzball solutions known so far are the JMaRT \cite{Jejjala:2005yu,Giusto:2007tt,AlAlawi:2009qe} and the running-Bolt \cite{Bena:2009qv, Bobev:2009kn} solutions. These are very special, non-generic solutions.} In this paper, we find evidence for a Penrose-type process for supertubes in the background of the non-extremal rotating Cvetic-Youm black hole, from the (static) analysis of probe supertubes.

There is also a second motivation for studying probe supertubes in non-extremal backgrounds. They are a step towards a better understanding of general, non-extremal multi-center black hole solutions. It is of importance to construct multi-center configurations, both for their intrinsic value as basic solutions in supergravity, as for the important role they play in the fuzzball proposal and its solution to the information loss problem (see \cite{Mathur:2005zp, Bena:2007kg, Mathur:2008nj, Balasubramanian:2008da, Skenderis:2008qn, Chowdhury:2010ct} for reviews and references). A lot of features can already be understood from treating one or more centers as probes. For instance, for supersymmetric multi-centered solutions, many interesting results were first observed in the probe limit. This led to the development of a rich literature on fully backreacted supersymmetric and non-supersymmetric extremal solutions \cite{Denef:2000nb}--\cite{Bossard:2011kz}  and new interesting phenomena.  The next natural line of investigation is  the status of multi-centered non-extremal solutions.   As a first step, one can treat one center as a probe in a multi-center background close to extremality. One expects that at least {\em meta}stable multi-center bound states should exist, as small deformation of the known BPS and non-BPS extremal multi-center solutions. For recent work along these lines see  \cite{Anninos:2011vn,Bena:2011fc}.\footnote{In \cite{Bena:2011fc}, supertube probes are considered in supersymmetric multi-center backgrounds with incompatible supersymmetries, such that the total configuration is non-extremal. The authors of \cite{Anninos:2011vn} treat  brane probes in a four-dimensional non-extremal D0-D4 black hole background.}

\medskip

We derive two main results in this paper. First, we prove that in a non-extremal Cvetic-Youm black hole background  probe supertubes can form both metastable and stable bound states with the black hole. It is the angular momentum interaction between supertube and background that provides the necessary repulsive force that allows for \emph{stable} bound states.  The Cvetic-Youm black hole can have two independent angular momenta; we find stable and metastable bound states both when the tube and the black hole rotate in the same plane and when they rotate in orthogonal planes. Second, in the D1-D5-P frame, we consider the D1-D5 near horizon limit. For small non-extremal excitations of the third charge (P), we again find metastable and stable bound states. We show that the stable bound states have a natural intepretation in the D1-D5 orbifold CFT, describing a new phase which is entropically dominant over the non-extremal black hole phase.

Finally, we discuss the organization of this paper. In section \ref{s:Background+Hamiltonian} we recall the non-extremal rotating three charge black hole solution in the M-theory frame and give the Hamiltonian of a probe supertube in this background. In section \ref{s:Minima} we find the minima of the potential. We 
show that the marginally stable bound states of the supersymmetric limit generically lift and become unstable. Near extremality however, there are both metastable and stable bound states. We study in detail the effect of energy above extremality and of the charges and angular momenta on the existence of the bound states.  In section \ref{s:Decoupling}, we consider the background in the D1-D5-P frame and analyze the supertube potential in the D1-D5 decoupling limit. We then analyze the phase diagrams in the AdS and the CFT and compare the results, with emphasis on the interpretation of the new phase. In section \ref{s:Discussion} we discuss our results and give future directions for research. There are several clarifying appendices. In appendix \ref{app:Conventions}, we give our notations and conventions, in appendix \ref{app:Hamiltonian} we derive the Hamiltonian in detail. In appendix \ref{section:D1D5CFT}, we give a short discussion of the  D1-D5 CFT and appendix \ref{app:Comparison_Denef} contains a comparison to the setup of \cite{Anninos:2011vn}.

\paragraph{Note:} After the main results of this work were obtained, the preprint \cite{Anninos:2011vn} appeared, whose results our similar to ours. Both results were found independently. In \cite{Anninos:2011vn}, four-dimensional static backgrounds are considered that lift to non-rotating black holes in five dimensions using the 4D/5D connection. A lot of qualitative features are similar. The differences are that we consider smooth probes  and allow for angular momenta; the nature of the repulsive forces responsible for the stable bound states differs in the two approaches. Also, our treatment makes it possible to map the results to the D1-D5 system.

\section{Supertubes in non-extremal black hole background}
\label{s:Background+Hamiltonian}
We want to discuss the physics of supertubes in the background of the five-dimensional non-extremal three charge Cvetic-Youm  black hole. First we discuss the background in the M-theory frame and then we give the Hamiltonian for a supertube in this background. The treatment of the minima of the Hamiltonian is deferred to section \ref{s:Minima}.

\subsection{Cvetic-Youm black hole in M-theory}
The Cvetic-Youm black hole \cite{Cvetic:1996xz,Cvetic:1996kv,Giusto:2004id} is a non-extremal, rotating three charge black hole of five dimensional supergravity. It has two angular momenta, in two independent planes in $\mathbb{R}^4$. We give the solution in the M-theory frame where it arises from a $T^6$ compactification. The three charges come from M2 branes wrapped on three orthogonal $T^2$'s inside $T^6$.

The solution depends on six parameters: a mass parameter $m$, three `boosts' $\delta_I$ related to the charges and angular momentum parameters $a_1,a_2$. The metric and gauge field are
\bea
ds_{11}^2 &=& -Z^{-2} H_m (dt + k)^2 + Z ds_4^2 + \sum_{I=1}^3 \frac{Z} {H_I}ds_I^2\,,\nn
A_3 &=& \sum_{I=1}^3 A^{(I)} \wedge \omega_I\,,\qquad A^{(I)} = \coth (\delta_I) H_I^{-1} (dt + k) + B^{(I)} - \coth (\delta_I) dt\,,\label{eq:11d_Background}
\eea
where  $Z = (H_1 H_2 H_3)^{1/3}$, $ds_I^2$ and $\omega_I$ are the flat metric and volume form on the $I^{\rm th}$ torus and the pure gauge term $\coth (\delta_I) dt$ in the gauge field was subtracted to make sure the electric potential dies off at spatial infinity. The rotation one-form $k$ and magnetic parts $B^{(I)}$ of the gauge fields are
\bea
k &=& \f{m}{f} \left[ -\frac{ c_1 c_2 c_3}{H_m} (a_1 \cos^2\theta\, d\psi + a_2\sin^2 \theta\, d\phi) + s_1s_2s_3 (a_2 \cos^2\theta\, d\psi + a_1\sin^2 \theta \,d\phi) \right]\,,\nn
B^{(I)}&=& \frac{m}{f H_m} \frac{ c_J c_K}{s_I} (a_1 \cos^2\theta \,d\psi + a_2\sin^2 \theta d\phi)\,,
\eea
with $I,J,K$ all different and we write
\be
c_I \equiv \cosh \delta_I\,, \qquad s_I \equiv \sinh \delta_I\,.
\ee
The solution is built from the functions
\be
H_I = 1 + \frac{m s_I^2}{f}\,,\qquad H_m = 1 - \frac{m}f\,, \qquad f = r^2 + a_1^2 \sin^2\theta + a_2^2 \cos^2 \theta\,.
\ee
The four-dimensional metric is 
\bea
ds_4^2 &=& \frac{f r^2}{g}dr^2 + f ( d \theta^2 + \sin^2\theta\, d \phi^2 + \cos^2 \theta\, d\psi^2)\nn
&&+ H_m^{-1} (a_1 \cos^2\theta \,d\psi + a_2\sin^2 \theta d\phi)^2 - (a_2 \cos^2\theta \,d\psi + a_1\sin^2 \theta \,d\phi)^2\,,\nn
g &=& (r^2 + a_1^2)(r^2 + a_2^2) - m r^2 \equiv (r^2 - r^2_+)(r^2 - r^2_-)\,.\label{eq:4d_Base}
\eea
The roots of the function $g(r)$ give the radial position of the inner and outer horizon of the black hole:
\be
(r_\pm)^2 = \frac 12 \left( m- {a_1^2}-{a_2^2} \pm \sqrt{\left(m- a_1^2-a_2^2\right)^2-4 a_1^2 a_2^2}\right)\,.
\ee 
In order for the surface $r^2 = r_+^2$ to be null and describe an event horizon, one needs to impose $m \geq (|a_1| + |a_2|)^2$, with equality for an extremal black hole \cite{Jejjala:2005yu}.

The ADM mass, electric charges and angular momenta of the black hole are 
\eal{
M_{ADM} &= \frac m 2 \sum_I \cosh 2 \delta_I\,, \qquad &J_\psi =-m( a_1 c_1 c_2 c_3 - a_2 s_1 s_2 s_3)\,,\\
\bQ_I &= \frac m 2 \sinh 2 \delta_I\,, &J_\phi =-m( a_2 c_1 c_2 c_3 - a_1 s_1 s_2 s_3)\,,
\label{eq:5dCharges}
}
where we have set $G_5 = \f{\pi}{4}$ as discussed in appendix \ref{app:Conventions}.

There are two extremal limits. The supersymmetric extremal limit is $m, a_1,a_2 \to 0$ and $|\delta_I| \to  \infty $  while keeping fixed the charges $\bQ_I$ and ratios $a_i/\sqrt{m}$. The four-dimensional base space becomes flat and one recovers the supersymmetric rotating three-charge BMPV black hole \cite{Breckenridge:1996is} with $M_{ADM} = \sum_I |\bQ_I|$. The non-supersymmetric extremal limit is obtained by putting $m = (|a_1| +|a_2|)^2$ and has $M_{ADM} > \sum_I |\bQ_I|$. 
 This is the `ergo-cold' black hole studied in \cite{Dias:2007nj}.

\subsection{Supertube Hamiltonian}

\begin{figure}[ht!]
\centering
\includegraphics[width=.35\textwidth]{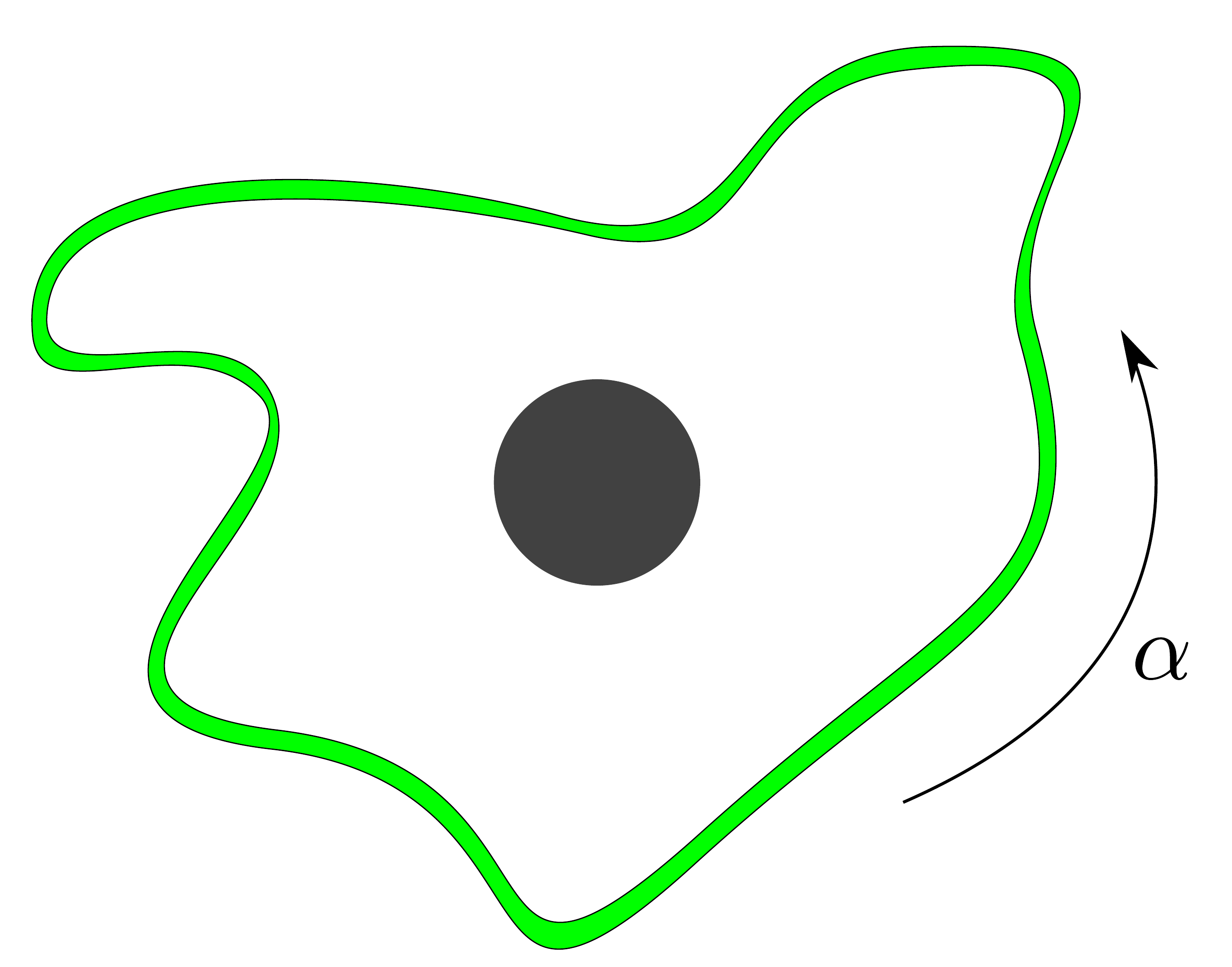}
\caption{\small Supertube embedded around a black hole.
\label{fig:SupertubeEmbedding}}
\end{figure}

We want to investigate the dynamics of supertubes in the non-extremal black hole background. 
A supertube is a brane configuration with two monopole brane charges, a dipole brane charge and angular momentum, that (locally) preserves eight supersymmetries \cite{Mateos:2001qs,Mateos:2001pi, Emparan:2001ux}. 
We consider supertubes with the two charges $\pQ_1$ and $\pQ_2$ corresponding to M2 branes on the first two $T^2$'s.\footnote{We use lower case for probe charges, upper case for background charges.}  The dipole charge, which we call $d_3$, is an M5 brane along those two $T^2$'s and along a one-cycle in the four-dimensional base which we parameterize by an angular coordinate $\a$ as shown in Figure \ref{fig:SupertubeEmbedding} and two constants $b_1,b_2$ describing its embedding as 
\be
\psi = b_1 \a\,, \quad\phi = b_2 \a\,.
\ee 
Because the M5 brane wraps a contractible cycle, the supertube carries no net M5 charge but an M5 dipole moment. The angular momentum of the tube is related to the other charges as
\be
j^{tube} = b_1 j^{\rm tube}_\psi +b_2 j^{\rm tube}_\phi = \frac{\pQ_1 \pQ_2}{d_3}\,.
\ee
The Hamiltonian for a supertube probe can be obtained from its DBI/WZ action. For details, see appendix \ref{app:Hamiltonian}, where the calculation is performed in a specific ten-dimensional duality frame. 

The Hamiltonian is
\bea
&&\calh =\nn
&&\f{1}{|d_3|} \frac{\sqrt{H_m H_1H_2H_3 g^{(4)}_{\a \a}}}{H_1H_2H_3 g^{(4)}_{\a \a} - H_m k_\a^2}\sqrt{\ti \pQ_1^2 + d_3^2 \frac{H_1H_2H_3 g^{(4)}_{\a \a} - H_m k_\a^2}{H_2^2}}\sqrt{\ti \pQ_2^2 + d_3^2 \frac{H_1H_2H_3 g^{(4)}_{\a \a} - H_m k_\a^2}{H_1^2}}\nonumber\\
&&+\f{1}{d_3} \frac{H_m k_\a}{H_1H_2H_3 g^{(4)}_{\a \a} - H_m k_\a^2} \ti \pQ_1 \ti \pQ_2 -\coth \d _1\, \f{ \tilde \pQ_1}{H_1}    - \coth \d _2 \,\f{ \tilde \pQ_2}{ H_2}- \f{1}{d_3} \coth \d _1 \coth \d _2 \frac{k_\a}{H_1 H_2}\nn
&&- \f m {d_3} \f {\bQ_3}{\bQ_1 \bQ_2} B^{(3)}_\a + \coth \d _1\ \pQ_1 + \coth \d _2\ \pQ_2 \,,
\label{eq:NonExtremalHamFullBulk}
\eea
where $k_\a,g^{(4)}_{\a \a},B^{(3)}_\a$ are the pullbacks of the rotation one form, the four-dimensional metric and the third magnetic field on the supertube worldvolume. The charges appear through the shifted quantities
\be
\ti \pQ_1 = \pQ_1 - d_3 A_\a^{(2)}\,,\qquad \ti \pQ_2 = \pQ_2 - d_3 A_\a^{(1)}\,, \label{ShiftedCharges}
\ee
where $A^{(I)}_\alpha$ are the pullbacks of the gauge fields on the supertube worldvolume.

We discuss two interesting limits of the background to understand the various terms. In the flat space limit, the second and third lines in \eqref{eq:NonExtremalHamFullBulk} vanish and the first line gives the Hamiltonian of a supertube in flat space $\calh_{flat} = 1/R \sqrt{\pQ_1^2 + R^2}\sqrt{\pQ_2^2 + R^2}$, where $R$ is the radius of the supertube. This term is like a position-dependent mass term. It stops the supertube from flying away to infinity due to the tension of the dipole M5 brane and also contains a centrifugal barrier for smaller radius. These two stabilize the supertube. 

In the supersymmetric limit, the first term in the last line drops out and one recovers a special case of  \cite{Bena:2011fc}, where the supertube Hamiltonian was computed in a general supersymmetric three-charge background. In this limit, $\coth \delta_1 = \coth \delta_2 = 1$ and one can prove that the Hamiltonian is always greater than the sum of the charges, with $\calh = \pQ_1 + \pQ_2$ only for supersymmetric minima. In particular, we recover the physics of supertubes in a BMPV background described previously in \cite{Bena:2004wt,Marolf:2005cx}. 

The only new term appearing in the Hamiltonian  in the non-extremal background  is the first one in the third line. This comes from the background magnetic gauge field. All other terms are the direct generalization of the corresponding terms in the supersymmetric BMPV background. We thus expect that at least close to supersymmetric extremality (small $m,a_1,a_2$; large $\delta_I$) we see similar physics as for the supersymmetric background. In particular, we anticipate the existence of (meta)stable bound states outside the black hole horizon. This intuition is confirmed in the next section.

\section{Minima of the potential and bound states}
\label{s:Minima}

In this section we wish to treat in depth the physics of supertubes in the Cvetic-Youm background. To that end we discuss the possible minima of the supertube potential. First, we concentrate on the supersymmetric limit and confirm results of \cite{Bena:2004wt,Marolf:2005cx} for supertubes in the BMPV background. In particular, we see that there is the possibility of having a supersymmetric configuration where the supertube settles outside the black hole horizon. Afterwards, we concentrate on the non-extremal background. We see that for the near-extremal black hole, depending on the orientation of the angular momentum of the background vs.\ that of the supertube, the configuration can either become stable, or metastable (having respectively lower or higher energy than when the tube falls to the horizon). Far away from extremality all minima of the Hamiltonian are lifted and the tube falls into the black hole. To get insight in the origin of the stable and metastable bound states, we discuss the effects of the various interactions (due to gravity, angular momenta, electric charges etc.) on general near-extremal configurations. We see that angular momentum is necessary in giving the repulsive force that accounts  for stable bound states.

To scan the effects of the various interactions on the existence of bound states in this section, we choose a `passive' point of view. We take positive supertube charges $\pQ_1,\pQ_2$ and only one unit of dipole charge viz.\ $d_3=1$.\footnote{Fixing $d_3 = 1$ does not affect the physics, as the Hamiltonian scales linearly in all the probe charges $\pQ_1,\pQ_2,d_3$ , see appendix \ref{app:Scaling}.} We also assume that the supertube  wraps the $\psi$ direction only while being in the $\theta=0$ plane 
\be
b_1 = 1\, , \qquad b_2 = 0\, , \qquad  \theta = 0\,,\label{eq:Embedding_Tube}
\ee
and that the angular momentum for the tube is positive. We thus have $J_\psi^{\rm tube} = \pQ_1 \pQ_2/d_3>0$ and $J^{\rm tube}_\phi =0$. In this section, we let the background charges and angular momentum vary (both in amplitude and orientation).

In all the plots in this section we choose the radial coordinate on the horizontal axis
\be
\rho^2 \equiv r^2 - r_+^2\,,
\ee
so it goes to zero at the horizon. The vertical axis measures the potential above its horizon value
\be
\tilde \calh(\rho) \equiv \calh(\rho) - \calh(\rho =0)\,.
\ee
Note that near-extremality, where most interesting physics happens, the probe approximation is only valid when $\pQ_1+\pQ_2 \ll M_{\rm ADM}$, so we restrict to relatively small tube charges.

\subsection{Supersymmetric background}

We consider the supersymmetric limit of the Cvetic-Youm black hole \eqref{eq:11d_Background}. The geometry becomes the five-dimensional rotating supersymmetric BMPV black hole \cite{Breckenridge:1996is} with  ADM mass $M_{ADM} = \bQ_1 + \bQ_2 + \bQ_3$ and self-dual angular momentum $J_\psi = -J_\phi \equiv J$ and the four-dimensional metric is just flat space.\footnote{We choose the supersymmetric limit with $m,a_1,a_2 \to 0^+,\,\delta_I \to +\infty$ while keeping $\bQ_I,m/\sqrt{a_i}$ fixed.}

The potential for a supertube in a generic supersymmetric three-charge background, of which the BMPV background is a special case, was recently discussed in \cite{Bena:2011fc}. Because of supersymmetry, everything can be described analytically and the probe approximation captures all the physics of the fully back-reacted solution \cite{Bena:2008dw}.

\begin{figure}[ht!]
\centering
\subfigure[Examples of bound states ($\pQ_1 \pQ_2 - \bQ_3 >0$).]{
\includegraphics[width=.45\textwidth]{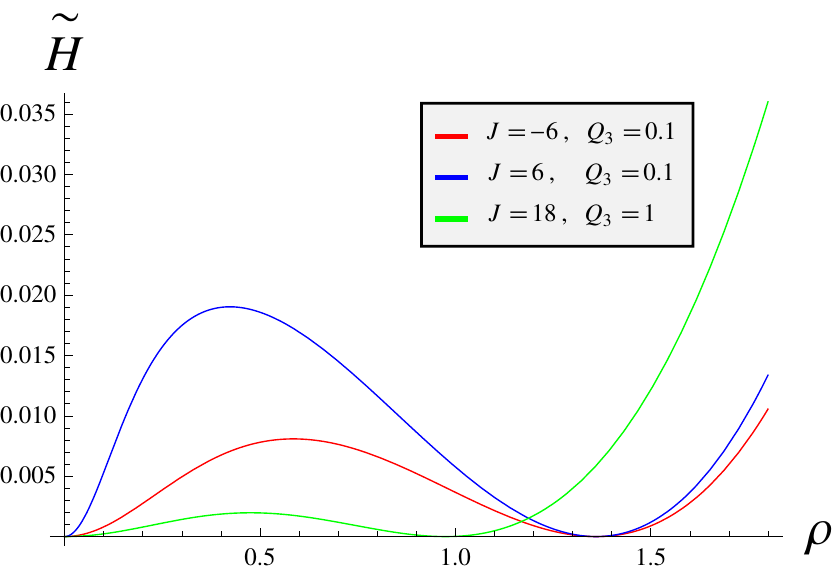}
\label{fig:BMPV_Hamiltonian-a}
}
\hspace{.02\textwidth}
\subfigure[No bound state ($\pQ_1 \pQ_2 - \bQ_3 < 0$).]{
\includegraphics[width=.45\textwidth]{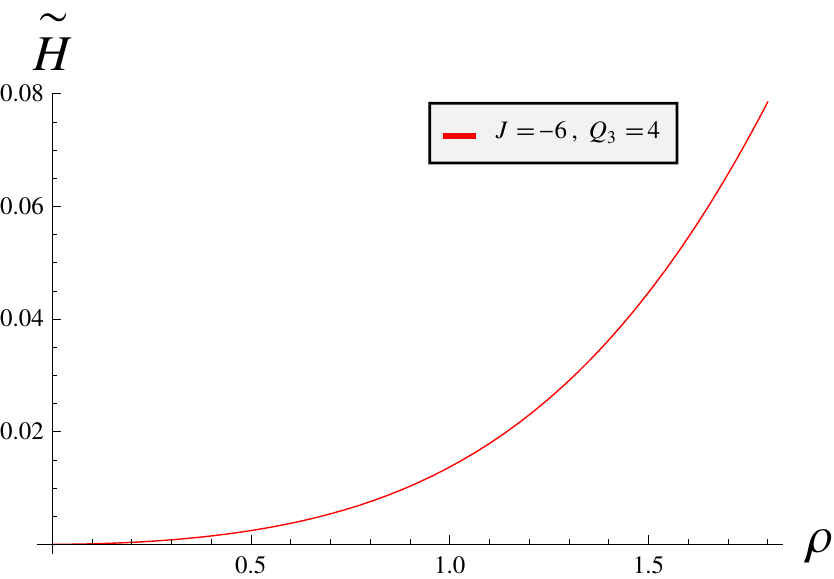}
}
\caption{The supertube Hamiltonian in the supersymmetric rotating black hole background. The tube wraps the  $\psi$-direction and has rotation in that plane only.  Furthermore, we have chosen the background and supertube charges   $(\bQ_1, \bQ_2) = (30,20)$, $(\pQ_1,\pQ_2) = (1.5,1.3)$. The third background charge $\bQ_3$ and the angular momentum $J$ are varied in the plots to show their effects.
On the left, examples of bound states. Varying $\bQ_3$ shifts the radial position of the minimum, while the relative orientation of background and tube angular momenta affects the height of the potential barrier (red vs.\ blue curve). The barrier is highest when they rotate in the same direction $(J_\psi>0)$. On the right we plot a representative case for which no bound state exists.
\label{fig:BMPV_Hamiltonian}}
\end{figure}

At the supersymmetric minimum, the potential is equal to the sum of the charges $\calh = \pQ_1 + \pQ_2$. For the BMPV background, one finds that the supersymmetric minimum of the tube potential is given implicitly as $\pQ_1 \pQ_2/d_3^2 = H_3 g_{\a \a}^{(4)}$ \cite{Bena:2008dw,Bena:2011fc}.  For the embedding \eqref{eq:Embedding_Tube} and $d_3=1$, this determines the supertube position $r = r_\star$ as 
\be
r_\star^2 = \pQ_1 \pQ_2 - \bQ_3\,.\label{eq:BMPV_Radius}
\ee
A non-trivial supersymmetric minimum only occurs when the right-hand side is positive. Note that the third charge $\bQ_3$ is the only background information in this equation: the values of the other charges $\bQ_1,\bQ_2$ and the angular momentum $J$ do not affect the position of the bound state. These other background parameters, however, do determine the height of the potential barrier, through the electrostatic and angular momentum interactions between the background and the tube. These features are illustrated in Figure \ref{fig:BMPV_Hamiltonian}.

From Figure \ref{fig:BMPV_Hamiltonian} we see that when the supertube angular momentum is aligned with that of the black hole ($J_\psi>0$) the barrier is higher than when they are opposite in direction ($J_\psi<0$). This suggests that like angular momenta repel while opposite angular momenta attract. However, the bound state with the supertube at $r = r_\star$ and the state where the black hole has merged with the supertube at $r = 0$ have the same energy independent of the sign of $J_\psi$. Thus from the point of view of the minima of the potential alone, we cannot answer any questions about stability.

\subsection{Effect of adding mass: Near extremal and  far from extremality}\label{subsec:CYBackGnd}

We consider the effects of going away from extremality. At fixed charges and angular momenta of the background, this is achieved by taking the parameter $m$ greater than zero:  
\begin{figure}[hb!]
\centering%
\subfigure[``Near''-extremal: $m = 0.01$.\label{fig:NearExtremal_Hamiltonian_a}]{
\includegraphics[width=.45\textwidth]{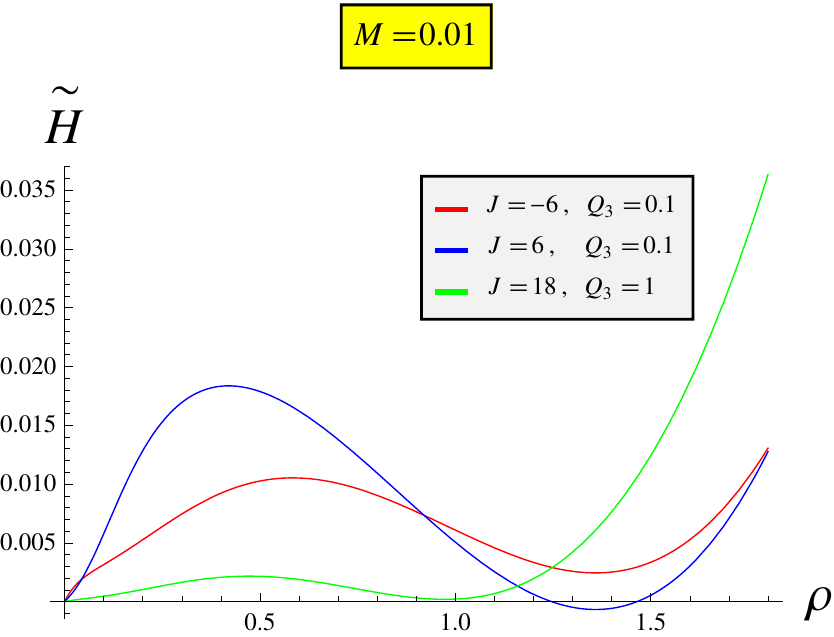}
}
\hspace{.05\textwidth}
\subfigure[Non-extremal: $m=0.9$.\label{fig:NearExtremal_Hamiltonian_b}]{
\includegraphics[width=.45\textwidth]{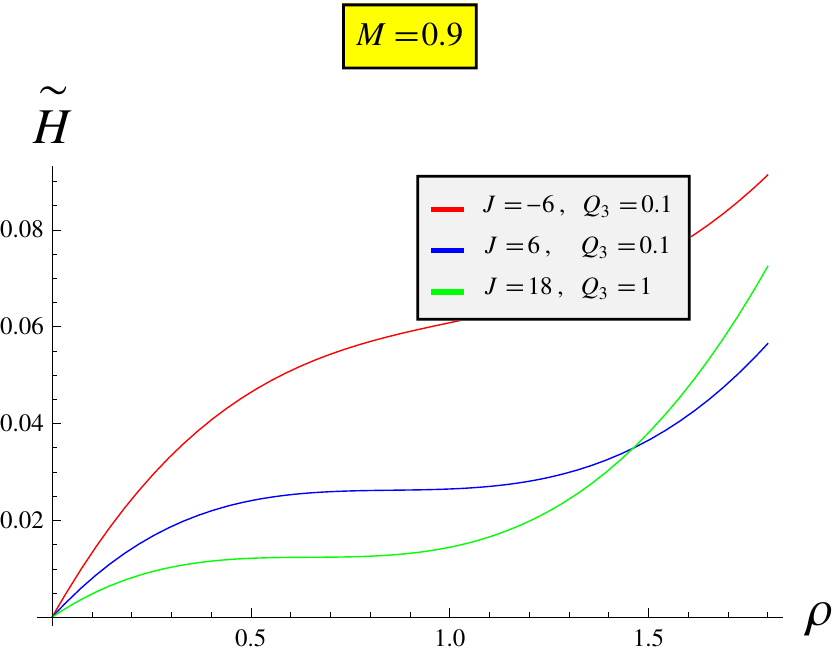}
}
\caption{The supertube Hamiltonian in the non-supersymmetric rotating black hole background.  The fixed  background charges are  $(\bQ_1, \bQ_2) = (30,20)$ and supertube charges  are  $(\pQ_1,\pQ_2) = (1.5,1.3)$. On the left, 3 choices of background and probe parameters near extremality ($m=0.01)$. On the right, the same system farther away from extremality ($m = 0.9)$. All plotted examples in this Figure have the same parameters as those in Figure \protect \ref{fig:BMPV_Hamiltonian-a}  (we again took self-dual angular momenta $J \equiv J_\psi = - J_\phi$); hence  Figure \protect \ref{fig:BMPV_Hamiltonian-a} gives the supersymmetric limit $(m=0)$ of these configurations. Near extremality, curves for supertubes corotating with the black hole ($J_\psi>0$) have their bound states becoming stable while curves for supertubes anti-rotating with the black hole ($J_\psi<0$) have their bound states becoming metastable. Far from extremality all the curves lift due to the dominating gravitational pull and there are no bound states.
\label{fig:NearExtremal_Hamiltonian}}
\end{figure}
Therefore, for larger $m$  the supertube probe will feel a stronger gravitational pull. We thus expect that for large $m$ the supertube potential will lift completely as the gravitational force dominates over all repulsive contributions. However for small $m$ the behavior is more interesting: we observe both metastable bound states (positive energy) and stable ones (negative energy).  The effect of increasing $m$ is illustrated in Figure \ref{fig:NearExtremal_Hamiltonian}. 

The charges and angular momenta of the examples in Figure \ref{fig:NearExtremal_Hamiltonian} are the same as those used for the supersymmetric BMPV limit in Figure \ref{fig:BMPV_Hamiltonian}; when $m=0$, we reproduce the latter figure. For small $m$ (Fig.\ \ref{fig:NearExtremal_Hamiltonian_a}), bound states persist. Again, like angular momenta of supertube and background ($J_\psi >0$) repel and opposite angular momenta ($J_\psi <0$) give an attractive force, but now they not only determine the height of the potenital barrier, but also the energy of the bound state. When the tube and the background rotate in opposite directions ($J_\psi<0$), the energy of the bound state is raised and it becomes metastable. When the tube and the black hole rotate in the same direction ($J_\psi >0$), the energy of the bound state is lowered and for $m$ small, there can be a stable state.  Raising $m$ further has the effect of lifting the energy of all the states, and far from extremality all bound states disappear from the spectrum, see Fig.\ \ref{fig:NearExtremal_Hamiltonian_b}.

\subsection{Effects of angular momentum}

We have seen that near extremality the supertube can form a metastable or stable bound state with the black hole. For the rest of this section we stay near extremality and investigate what the effects of the angular momentum-induced interactions are on the dynamics of the probe supertube. 

First we prove that when the background has no rotation ($J_\psi = J_\phi=0$), there are no stable bound states possible. We then show, in examples, that there are again metastable and stable bound states when angular momentum is turned on, both when the supertube rotates in  the same plane as the black hole and when it rotates in the orthogonal plane.

\subsubsection{No angular momentum, no stable bound states}

We prove that when the background has $J_\phi = J_\psi =0$, the minimum of the potential has a higher energy than at the horizon:
\be
\tilde \calh \equiv \calh - \calh \big |_{\rho = 0} >0\,,
\ee
and there can only be metastable bound states.

The Hamiltonian \eqref{eq:NonExtremalHamFullBulk} in the non-rotating CY background is	
\be
\tilde  \calh \equiv \calh - \calh|_{\rho =0} = \sqrt{H_m} (\calh_1 + \calh_2)\,,
\ee
with the definitions
\bea
\calh_1 &=& {\sqrt{d_3^2 \frac{R^2}{H_1^2 H_2^2}+\frac 1 {d_3^2}\frac{\pQ_1 ^2 \pQ_2^2}{ R^2}  + \pQ_1^2 + \pQ_2^2}}\,,\nn
\calh_2 &=& - \sqrt{H_m}\left(\tanh \d_1 \frac{\pQ_1}{H_1}+ \tanh \d_2 \frac{\pQ_2}{H_2}\right)\,,\nn
R^2 &=& H_1 H_2 H_3 g_{\a\a}^{(4)}\,.
\eea
It is enough to focus on the case where the electric charges of the supertube and the background have the same orientation ($\delta_I \pQ_I>0$) because then the electric repulsion is strongest (both terms in $\calh_2$ give a negative contribution). 

Consider the combination:
\be
\bar \calh_2 =- \left( \frac{\pQ_1}{H_1}+  \frac{\pQ_2}{H_2}\right)\,.
\ee
We find
\be
(\calh_1)^2 - (\bar \calh_2)^2 = \left(\frac{\pQ_1 \pQ_2}{R} -\frac{R}{H_1 H_2}\right)^2\,.
\ee
Since $\calh_1 >0$, it follows that also
\be
\calh_1 + \bar \calh_2 \geq 0\,.\label{eq:SUY-like_Bound}
\ee
Furthermore, because $H_m <1, |\tanh \delta_I| <1$, we have that $|\calh_2| < \bar \calh_2 $. Combined with \eqref{eq:SUY-like_Bound}, this gives that 
\be
\tilde \calh = \sqrt{H_m}(\calh_1 + \calh_2) > 0\,. 
\ee

Equality can only be obtained in the supersymmetric limit ($H_m=1,|\tanh \delta_I|=1$):
\be
\tilde \calh_{\rm susy}   = \calh_1^{\rm susy} +\calh_2^{\rm susy} \geq 0 \,,\label{eq:SUSY_Bound}
\ee
with equality when $\frac{\pQ_1 \pQ_2}{R} - \frac{R}{H_1 H_2} = 0$, or $\pQ_1 \pQ_2 = H_3 g_{\a\a}^{(4)}$, the susy supertube radius relation.

In conclusion, we see that one needs a non-zero background angular momentum to allow for stable black hole--supertube bound states (with $\ti \calh<0$).

\subsubsection{Black hole and supertube rotating in the same plane}

\begin{figure}
 \subfigure[Co-rotating\label{fig:tube-same-plane_a}]{
\includegraphics[width=.44\textwidth]{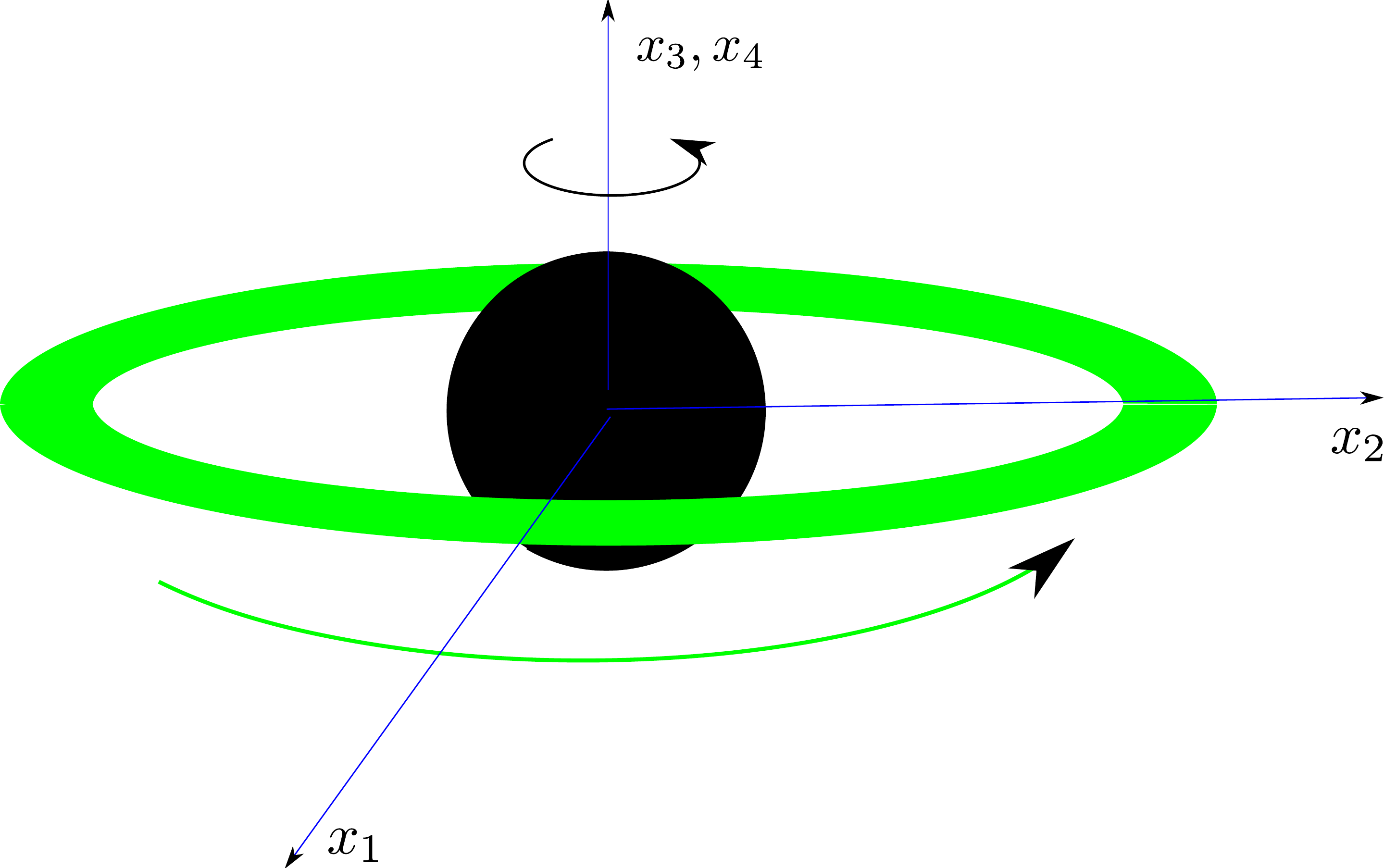}
}
\hspace{.04\textwidth}
\subfigure[Anti-rotating\label{fig:tube-same-plane_b}]{
\includegraphics[width=.44\textwidth]{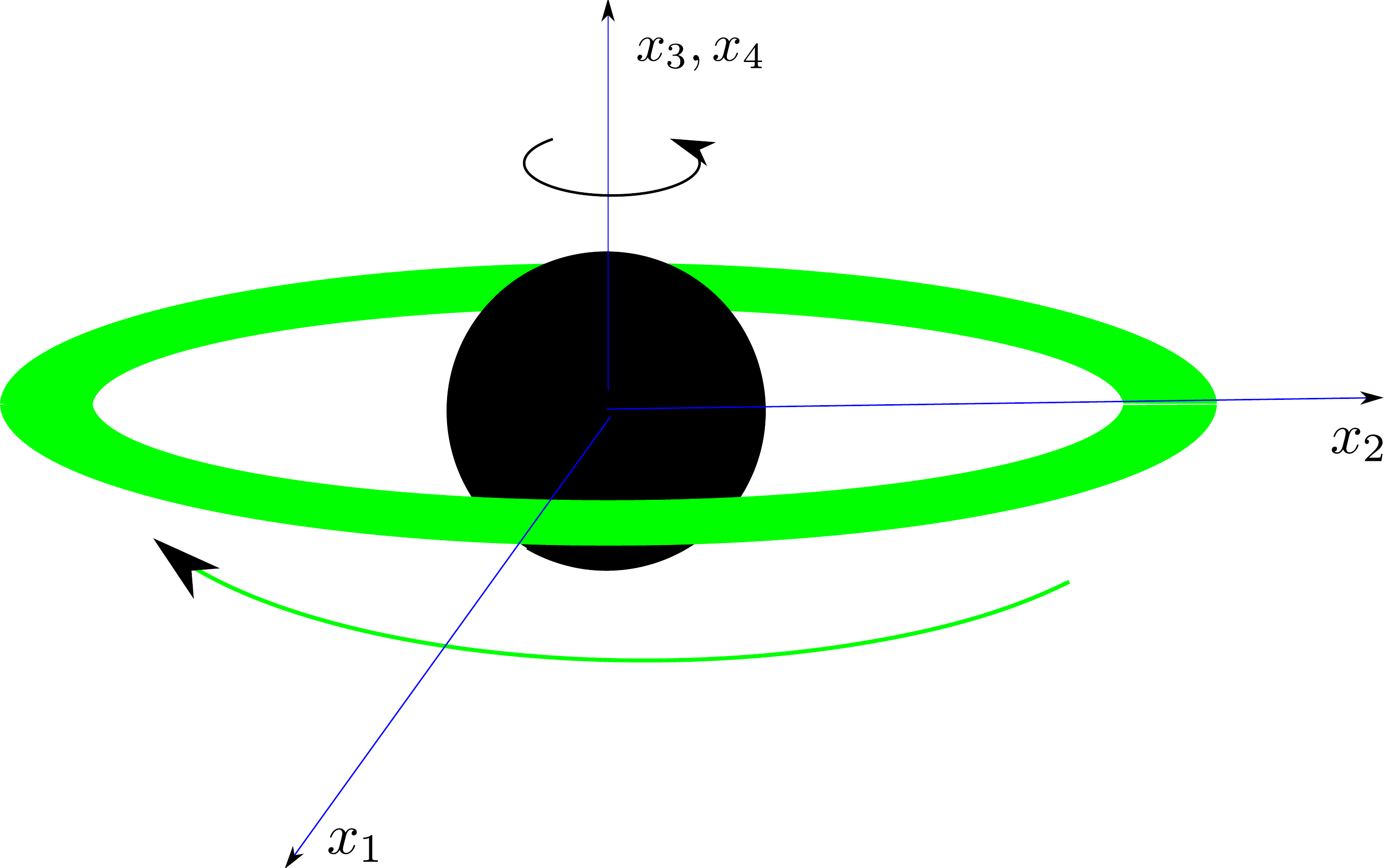}  
}
\caption{\sl The black hole and the supertube rotating in the same plane. The four-dimensional space is spanned by four cartesian coordinates $x_1,x_2,x_3,x_4$. In this figure, the black hole has rotation in the $x_1,x_2$ plane and so does the supertube. In (a) the tube rotates in the same direction as the black hole while in (b) the tube rotates in the opposite direction to the black hole.
\label{fig:tube-same-plane}}
\end{figure}

\begin{figure}[ht!]
\centering
\subfigure[Rotation in same plane ($J_\phi=0$).\label{fig:J_SamePlane}]{
\includegraphics[width=.45\textwidth]{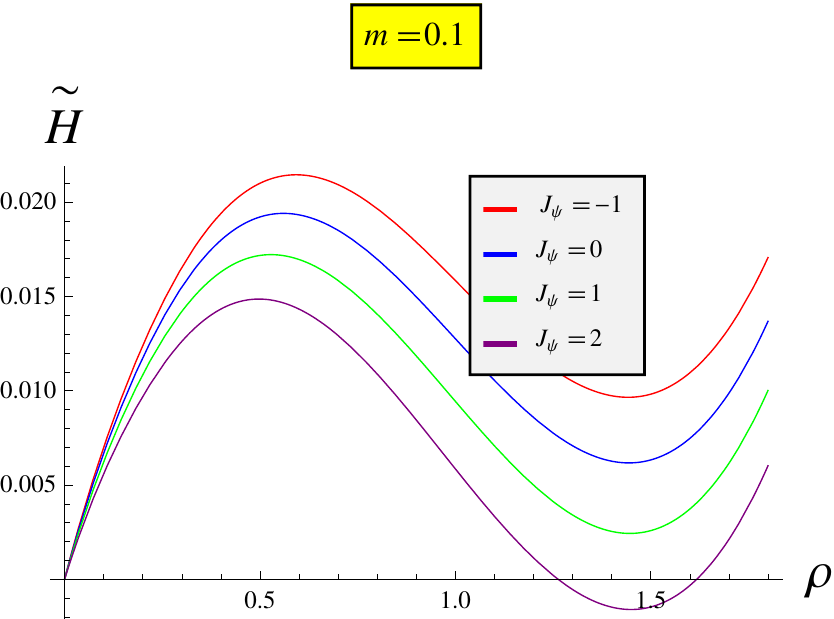}
}
\hspace{.05\textwidth}
\subfigure[Rotation in orthogonal planes ($J_\psi=0$).\label{fig:J_OtherPlane}]{
\includegraphics[width=.45\textwidth]{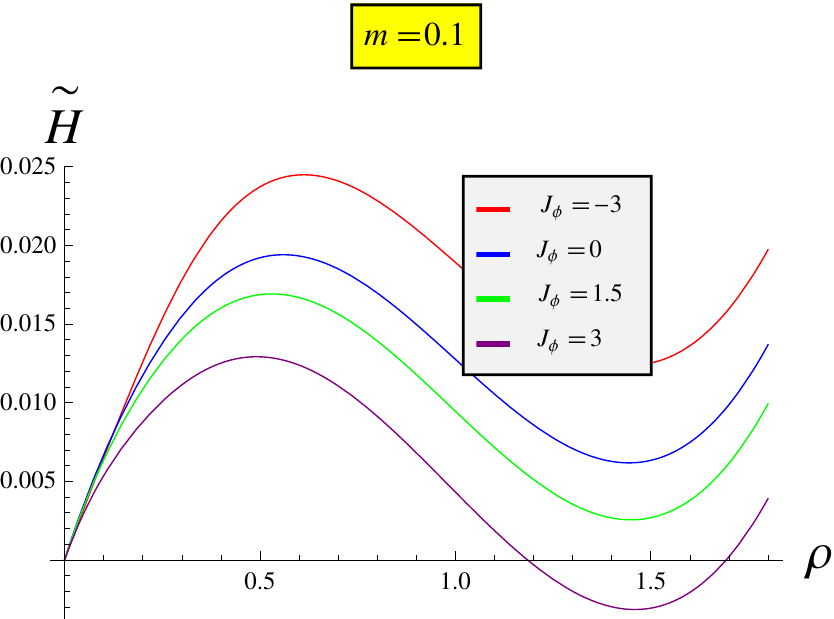}
}
\caption{Effects of relative orientation of angular momenta on the probe potential. The background has charges $(\bQ_1,\bQ_2,\bQ_3) = (20,20,0.1)$. The tube has charges $(\pQ_1, \pQ_2) = (1.5,1.5)$. On the left, $J_{\phi}=0$; tube and background rotate in the same plane. This plot shows that increasing $J_\psi$ lowers the energy of the minimum. For sufficiently large $J_\psi$ there are stable bound states. On the right $J_\psi =0$; the supertube and black hole rotate in orthogonal planes. Again the potential, and the value of the mininum, go down with increasing $J_\phi$.\label{fig:J_Planes}}
\end{figure}

We consider a background with $J_\phi=0$. Since we took $J^{\rm tube}_\phi =0$, this means that the supertube and the black hole rotate in the same plane. See Figure \ref{fig:tube-same-plane} for a pictorial representation. 
The strength of the interaction between the angular momenta of the background and the tube determines the stability of the bound state when we keep all other charges fixed. As we noticed before, the relative sign of $J_\psi$ and the tube angular momentum  is important. By choosing the same orientation and $J_\psi$ large enough, the bound state can become stable, see Figure \ref{fig:J_SamePlane}. 

\subsubsection{Black hole and supertube rotating in orthogonal planes}

When $J_\psi =0$, the supertube rotates in a plane orthogonal to the rotational plane of the black hole. Again, the sign and magnitude of $J_\phi$ determines (meta)stability of the  bound state. We observe that when $J_\phi<0$ and large, the bound state becomes stable.

This might come as as suprise, as one would expect that, since the tube and black hole rotate in orthogonal planes, the sign of $J_\phi$ would not matter. The reason of the seeming inconsistency is that $J_\phi \to -J_\phi$ is not a symmetry of the  Cvetic-Youm background. Rather, one should consider the discrete transformation $(J_\phi,\bQ_3) \to (-J_\phi,-\bQ_3)$ (see section \ref{app:Scaling}), which leaves the supertube Hamiltonian invariant.  In this sense, we can restrict to $J_\phi>0$ and attribute the origin of the bound states to a combined effect in $\bQ_3$ and $J_\phi$. We come back to this below.
\begin{figure}
\subfigure[Projection on $x_1,x_2$ plane]{
\includegraphics[width=.36\textwidth]{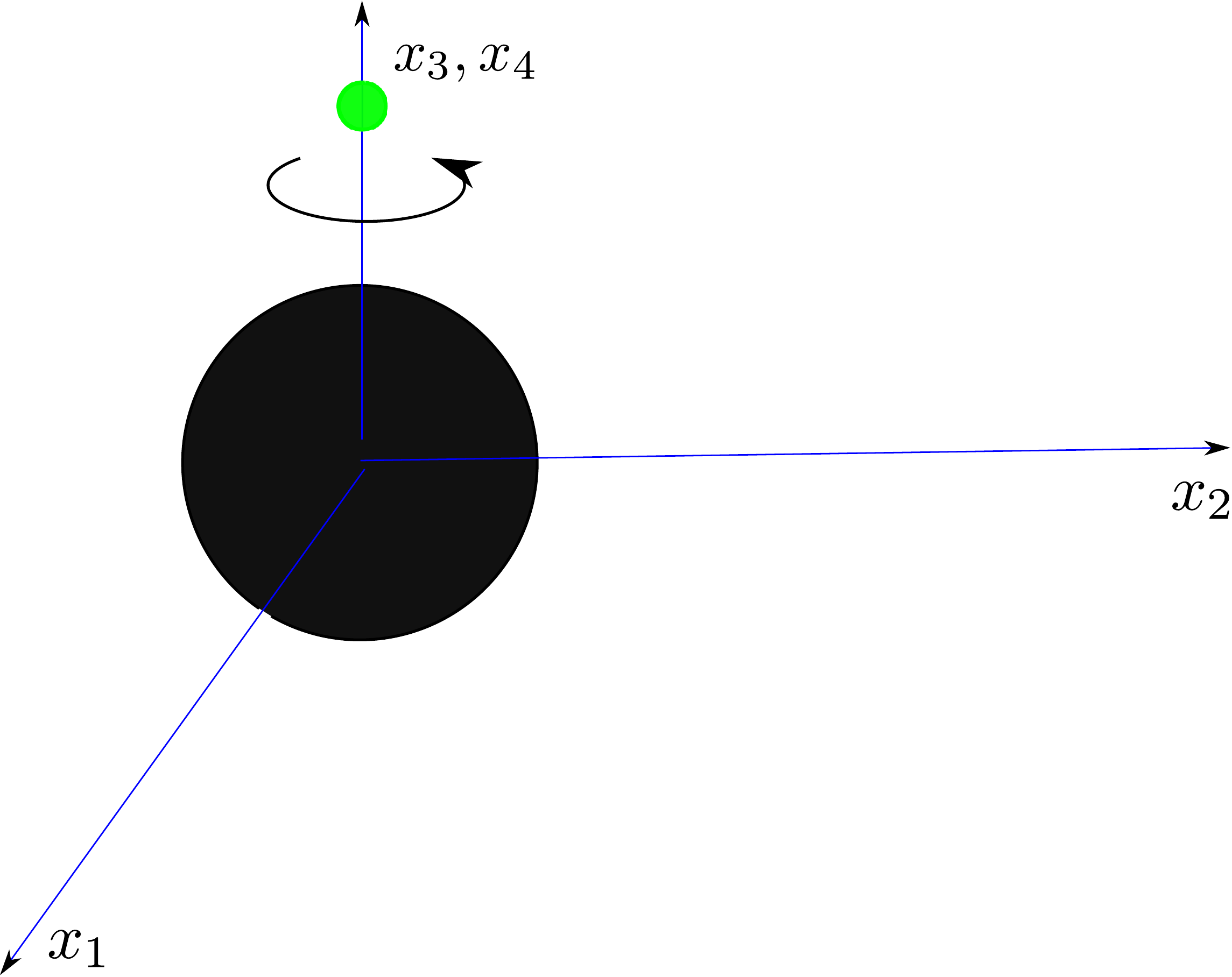}
}
\hspace{.1\textwidth}
\subfigure[Projection on $x_3,x_4$ plane]{
\includegraphics[width=.45\textwidth]{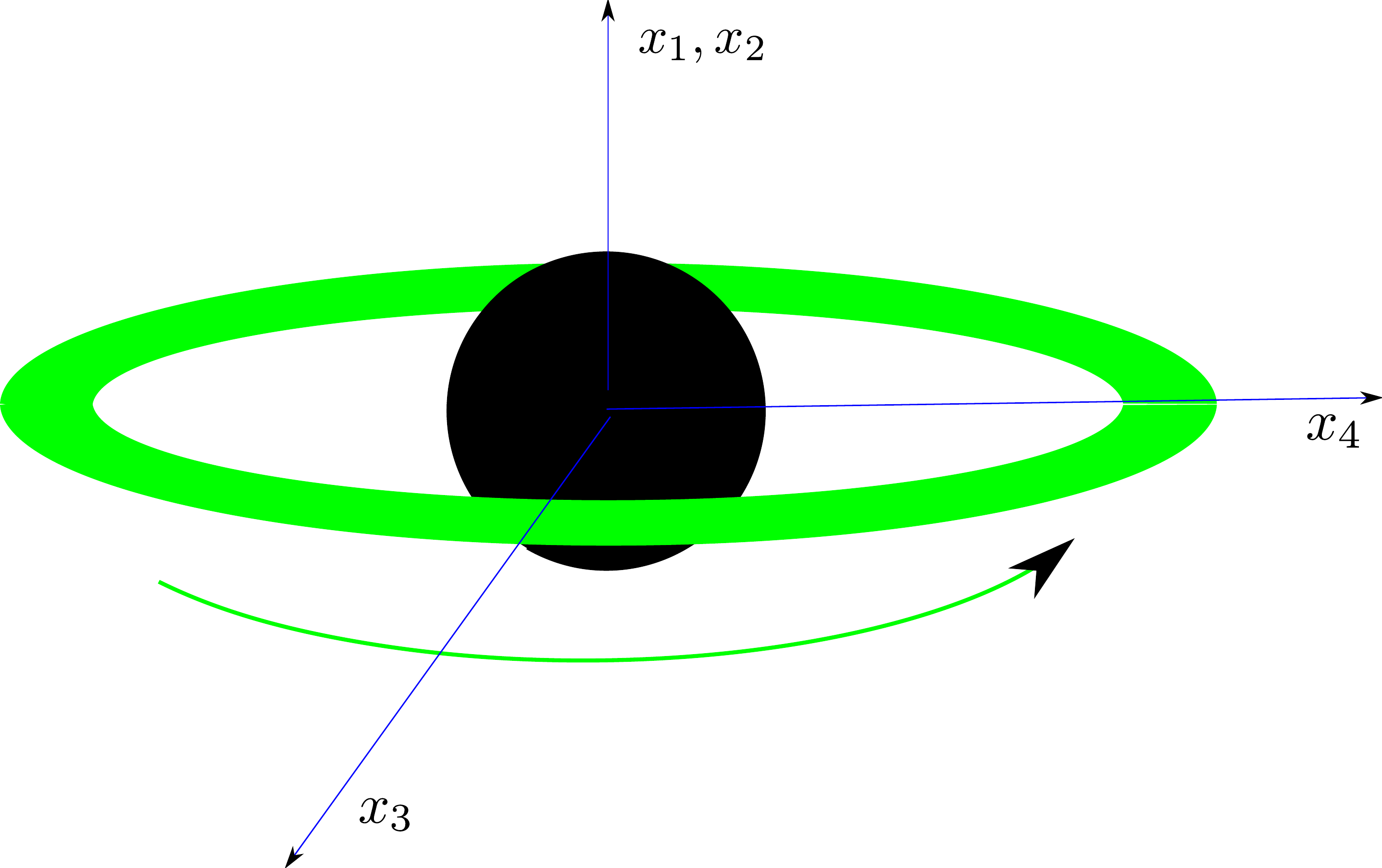}  
}
\caption{\sl We visualize a black hole with rotation in the $x_1,x_2$ plane and a supertube rotating in the $x_3,x_4$ plane. On the left, we project on the plane of rotation of the black hole; the supertube is a point on the vertical axis which represents the orthogonal plane. On the right, we project on the plane of rotation of the supertube. In both figures the black hole is a pointlike obect in the origin.
\label{fig:tube-orthogonal-plane}}
\end{figure}

\subsection{Effects of electric charges}

Here we study the effect of varying the electric charges on the existence and stability of bound states. Since the tube only has electric charges $\pQ_1,\pQ_2$ along the first two tori  we consider the two physically distinct possibilities: varying the third background charge $\bQ_3$, and varying the background charges $\bQ_1,\bQ_2$.

\begin{figure}[ht!]
\centering
\subfigure[Changing $|\bQ_3|$ with $J_\phi=0$. \label{fig:Q2_Effects_CoRotating}]{
\includegraphics[width=.45\textwidth]{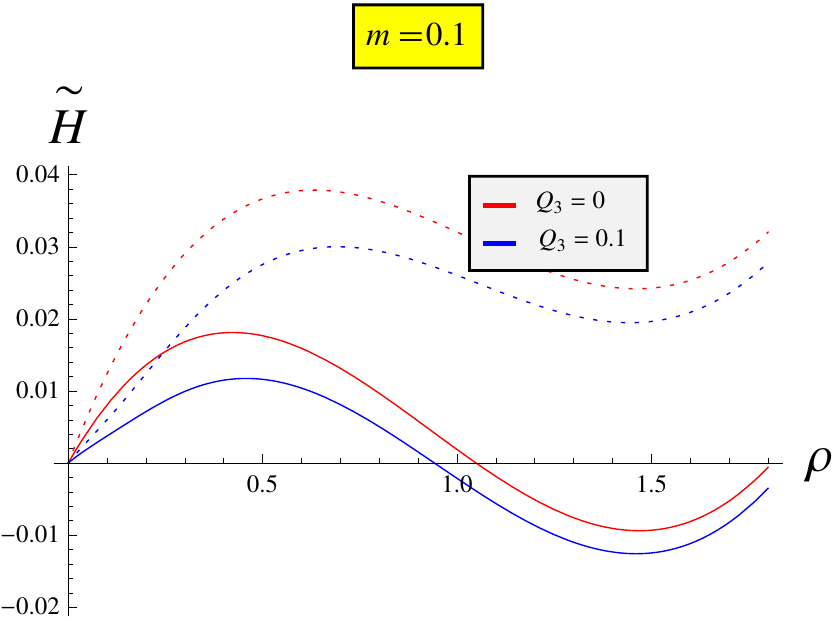}
}
\hspace{.05\textwidth}
\subfigure[Changing $\bQ_3$ with $J_\psi=0$. \label{fig:Q2_Effects_OrthogonalRotating}]{
\includegraphics[width=.45\textwidth]{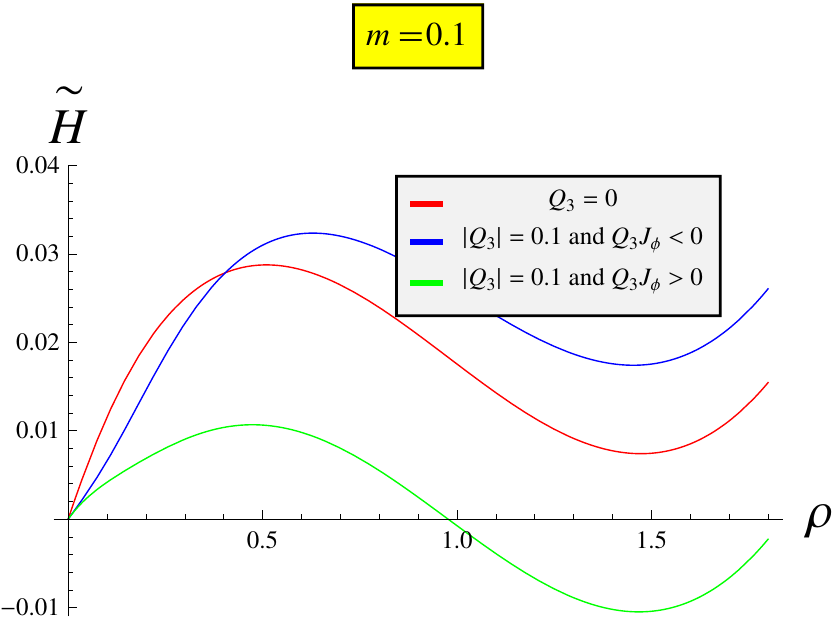}
}
\caption{Effects on the supertube potential of changing the third background charge. The fixed background charges are $(\bQ_1, \bQ_2)=(20,20)$ and the supertube charges are  $(\pQ_1, \pQ_2) = (1.5,1.5)$. The first figure is with $J_\phi=0$ and $|J_\psi|=3$ (i.e. with the supertube and the black hole rotating in the same plane). The dashed curves have $J_\psi<0$, the connected curves $J_\psi>0$. The effect of increasing $|\bQ_3|$ is to lower the minima. The second figure is with $J_\psi=0$ and $|J_\phi| = 3$ (i.e. with the supertube and the black hole rotating in orthogonal planes). When $\bQ_3 J_\phi<0$ the minimum is raised while when   $\bQ_3 J_\phi>0$ the minimum is lowered.}
\end{figure}

We first vary $\bQ_3$.  When the supertube rotation is in the same plane as the black hole's ($J_\phi=0$), we have the results in Figure \ref{fig:Q2_Effects_CoRotating}. This situation has a $\bQ_3 \to -\bQ_3$ symmetry as we would expect. The effect of increasing $|\bQ_3|$ in the case of both stable and metastable states is to lower the curve and therefore the minima, and to shift the radial position of the minimum towards the horizon.

The results of changing the third charge when the tube angular momentum is orthogonal to that of the black hole ($J_\psi=0$) are plotted in   \ref{fig:Q2_Effects_OrthogonalRotating}. As discussed before, the potential has a $(\bQ_3, J_\phi) \to (- \bQ_3, -J_\phi)$ symmetry. We see that when $ \bQ_3 J_\phi <0$ then increasing $|\bQ_3|$ raises the minima while when $ \bQ_3 J_\phi >0$ then increasing $|\bQ_3|$ lowers the minima. There can only be stable bound states for $ \bQ_3 J_\phi >0$, $|\bQ_3 J_\phi|$ large enough.

\begin{figure}[ht!]
\centering
\subfigure[Changing $\bQ_2$ with $|J_\psi=2|$.\label{fig:Q_Effects_12}]{
\includegraphics[width=.45\textwidth]{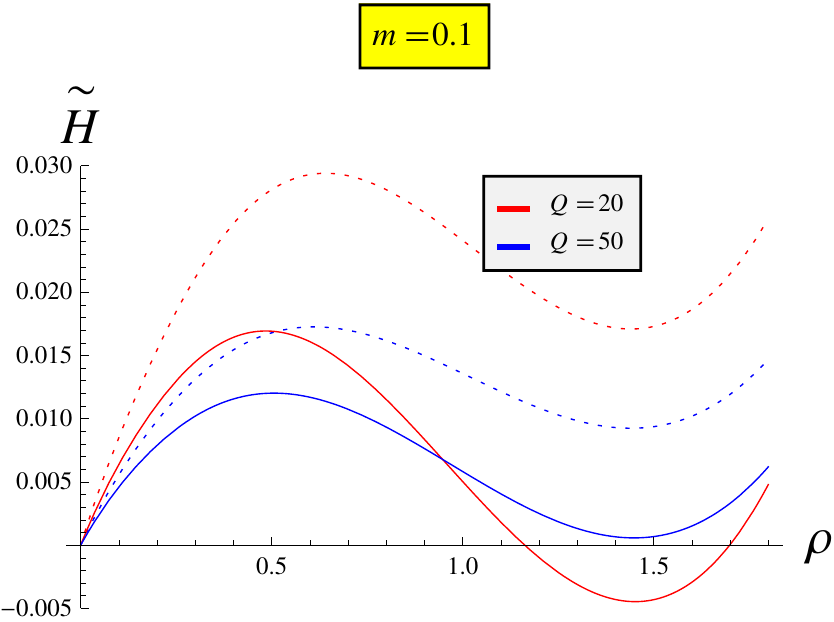}
}
\hspace{.05\textwidth}
\subfigure[Changing $\bQ_1$ and $\bQ_2$.\label{fig:Q1Q2Negative}]{
\includegraphics[width=.45\textwidth]{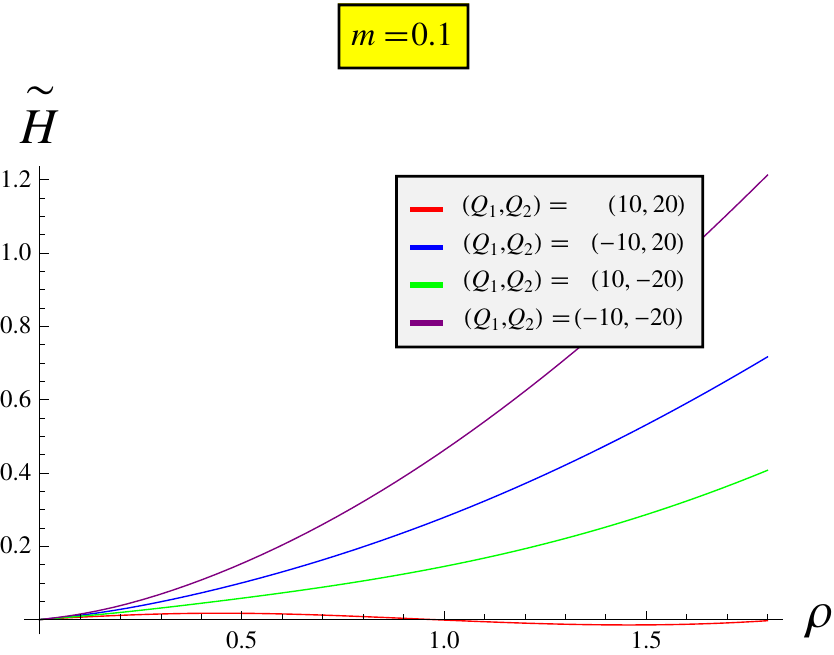}
}
\caption{Effects on the supertube potential of changing the first and second background charge. In both cases, the charges of the probe are  $(\pQ_1,\pQ_2)=(1.5,1.5)$. The figure on the left shows the effect of changing $Q_2$. It has fixed charge $(\bQ_1,\bQ_3,|J_\psi|,J_\phi) = (20,0.1,2,0)$ The dotted lines have $J_\psi<0$, the connected lines $J_\psi>0$. We see the effect of increasing $\bQ_2$ in both cases is to reduce the overall {\em amplitude} of the potential, leading to bound states only when tube and black hole rotate in the same direction ($J_\psi >0$).
The figure on the right has fixed background charges are $(\bQ_3, J_\psi, J_\phi) =(.1, 2, 0)$ \label{fig:Q2Postitive}. It shows the effect of the relative orientation of the tube and background charges. We see that a stable state for $\bQ_1>0,\bQ_2>0$ is no longer a bound state when at least one of these charges changes sign.}
\end{figure}

Finally, we study the impact of the background charges $\bQ_1,\bQ_2$ on the supertube potential. There are two effects at play. When ($\bQ_1,\bQ_2$)  have the same sign as respectively ($\pQ_1,\pQ_2$) then increasing   the magnitude of the background charges $\bQ_1$ or $\bQ_2$, reduces the overall amplitude of the potential. This is shown by varying $\bQ_2$ in Figure \ref{fig:Q_Effects_12}. Again, we only find bound states when the angular momenta of tube and black hole are aligned; they only appear for $\bQ_2$ small enough. The second effect has to do with the relative orientation of the charges of the black hole and those of the supertube, see Figure \ref{fig:Q1Q2Negative}. When ($\bQ_1,\bQ_2$)  have the same sign as respectively ($\pQ_1,\pQ_2$), the electrostatic repulsion is maximal and bound states can appear. When one or both of the pairs $(\bQ_1,\pQ_1),(\bQ_2,\pQ_2)$ have opposite orientation, attractive forces dominate and there are no bound states.

\subsection{Conclusions}
\label{ss:Conclusions_Sect3}

We find that near-extremality ($m$ small), bound states where the supertube settles at a certain radius from the black hole, are possible. Because of the large parameter space,\footnote{There are three background charges $\bQ_I$, two independent angular momenta $J_\psi,J_\phi$, energy above extremality, proportional to $m$, and three independent probe charges $d_3,\pQ_1,\pQ_2$. There are two continuous scaling symmetries (section \ref{app:Scaling}), such that we effectively have a 7-dimensional parameter space.} we have focused on a qualitative treatment of these bound states. With various plots, we have scanned the parameter space and found that for the bound states to exist, the electric charges of the probe $\pQ_1,\pQ_2$  need  to be oriented along $\bQ_1,\bQ_2$ of the background, and one needs $\bQ_3$ small enough. Also, to a large extent it is $\bQ_3$ which determines the radial position of the supertube. These features are similar to those for supertubes in the supersymmetric BMPV limit of the background.

New features compared to the supersymmetric limit are that near extremality a bound state can become metastable (positive energy) or stable (negative energy), where we normalized the energy to zero at the horizon. Whether a bound state is stable or metastable, depends mainly on the angular momentum interaction. We considered black holes with rotation in one plane. For a supertube rotating in the same plane as the black hole, the orientation determines stability. When the angular momenta are anti-aligned, there can be only metastable states. When they point in the same direction, the repulsive centrifugal force brings down the energy of the minimum and for large enough background angular momentum, the bound state can become stable. For the supertube and the black hole rotating in orthogonal planes, also the interaction with $\bQ_3$ plays a role, on top of the angular momentum $J$ of the black hole. When $\bQ_3 J <0$, we have only metastable bound states, when $\bQ_3 J>0$, stable states are possible when this combination is large enough.

\medskip

When a metastable supertube--black hole bound state exists, it is important to understand what the configuration where the tube sits at the black hole horizon  corresponds to (`merger' of the supertube and the black hole). The merger describes the end state of the decay of the bound state and should be a well-defined (black) object. In particular, pushing a metastable supertube in the black hole should not make it possible to overspin the black hole. Similarly, when a stable supertube--black hole bound state exists, it is important to understand when it is the final state of a process where a supertube pops out of a rotating black hole and when it cannot be understood in this way. These questions require a better understanding of the phase diagrams; we leave them to future work.

\section{Decoupling limit}
\label{s:Decoupling}

In this section, we turn to a frame where the three background charges describe a D1-D5-P system. We find a thermodynamic instability in the D1-D5 CFT along the lines of the one found in \cite{Bena:2011zw}.  Unlike \cite{Bena:2011zw}, we consider a non-supersymmetric instability. We also give the supertube potential in the appropriate decoupling limit of the Cvetic-Youm  geometry that is dual to the D1-D5 CFT and we find that the rich physics of metastable and stable bound states survives the decoupling limit.  The main purpose of this section is to discuss in detail the regions in parameter space where (meta)stable bound states exist in the D1-D5 CFT and in the bulk and to show how the thermodynamic instability in the CFT maps to the existence of stable bound states in the bulk, signaling a dynamic instability of the black hole.

\subsection{Background and motivation}

\subsubsection{Motivation}

In \cite{Bena:2011zw} a supersymmetric thermodynamic instability in the bulk was mapped to a supersymmetric thermodynamic instability in the dual CFT. The setup in  \cite{Bena:2011zw} was similar to the one being studied in the current paper but in addition it was supersymmetric. Supersymmetry allowed the study of fully back-reacted supertubes in the presence of BMPV black holes,\footnote{The BMPV black hole  \cite{Breckenridge:1996is} is the supersymmetric limit of the non-extremal Cvetic-Youm black hole.}  using the known construction of multi-center BPS solutions.  

\begin{figure}[tbh]
\subfigure[Phases in the bulk.\label{fig:MoultingBHs_a}]{
\epsfxsize=7cm \epsfbox{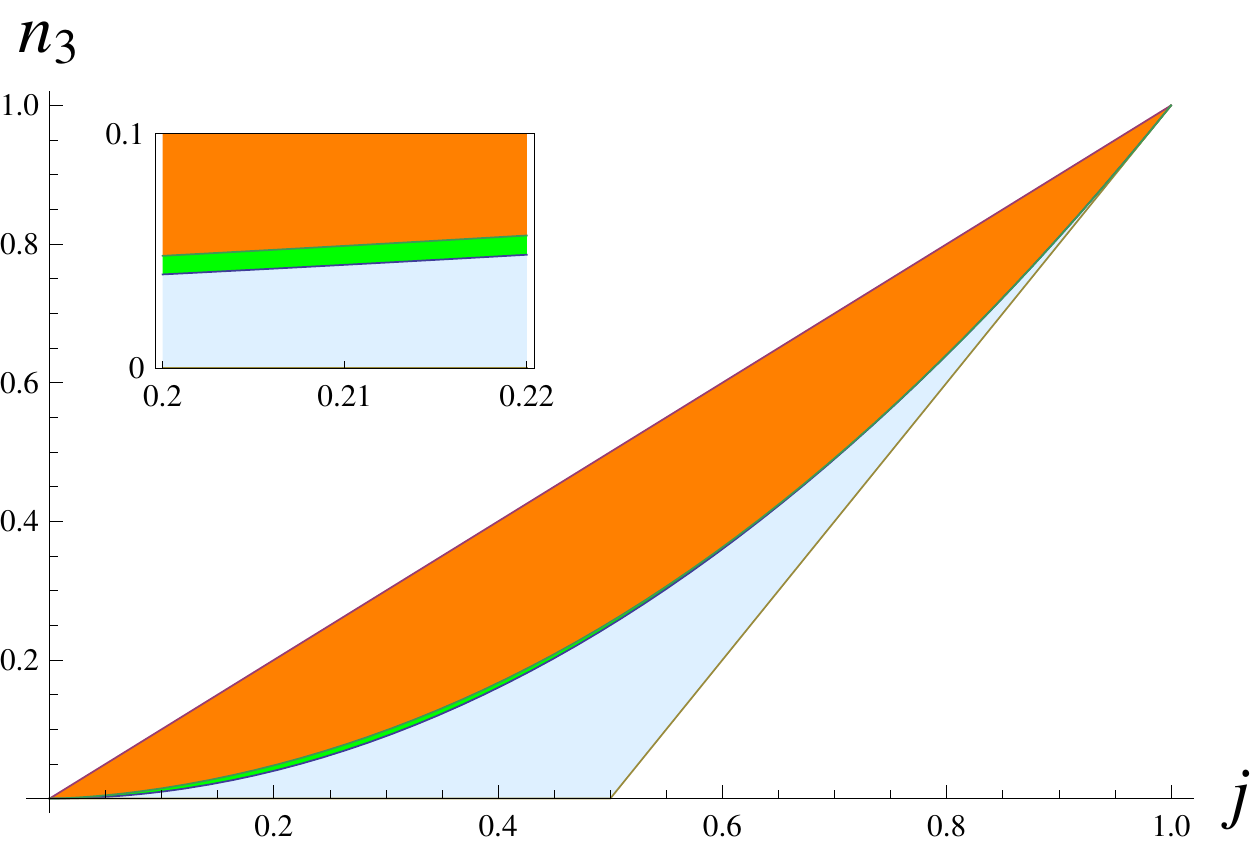}
}
 \hspace*{.3cm}
\subfigure[Phases in the CFT.\label{fig:MoultingBHs_b}]{
\epsfxsize=7cm \epsfbox{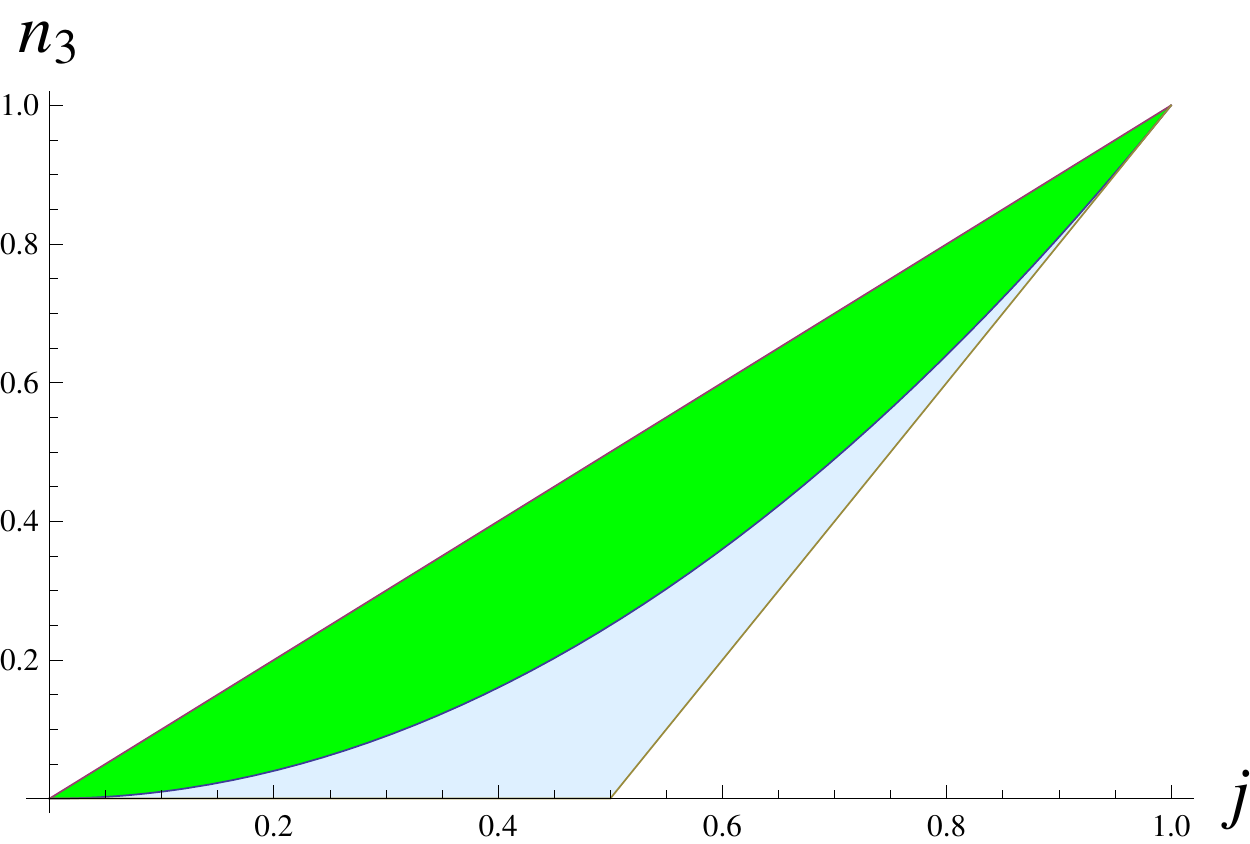}
}
\caption{\sl Existence and entropy of the new phase found in \cite{Bena:2011zw} and the single BMPV phase in the \mbox{$(j,n_3)$--plane}, where $j$ is proportional to the angular momentum of the black hole and  $n_3$ to its electric charge $\bQ_3$, see eq.\ \protect \eqref{eq:IntensiveQuantities} below.  In the bulk (left) the  supertube and BMPV phase can co-exist with the BMPV in the orange region but has smaller entropy. In the green region the supertube and BMPV is entropically dominant over the BMPV phase. In the blue region the bound state can exist even though the single BMPV itself cannot. In the CFT (right) the long string and short string condensate phase exists and dominates entropically over the long string phase in the green region. In the blue region the long string phase does not exist but the long string and short string condensate phase continues to exist.
  \label{fig:MoultingBHs}}

\end{figure}

It was found that in the bulk, supertubes can coexist with the BMPV black hole near the cosmic censorship bound $|\bQ_1 \bQ_2 \bQ_3 |- J^2 \geq 0$.\footnote{Absence of naked closed causal curves requires $|\bQ_1 \bQ_2 \bQ_3| - J^2 \geq 0$.}  In a small region, very close to this bound, the bound state of supertube and BMPV is entropically dominant over just a BMPV. This bound state also exists for charges that do not obey the cosmic censorship bound (and for which there is no BMPV black hole), all the way to the unitarity bound, which is a bound on the maximum rotation for a given mass coming from the dual D1-D5 CFT. This is shown in Figure \ref{fig:MoultingBHs_a}. In the dual CFT this thermodynamic instability manifested itself as the long string sector being entropically dominant  far away from the cosmic censorship bound while a sector with one long string and a short string condensate comes into existence and dominates entropically close to the  cosmic censorship bound.\footnote{See section \ref{sss:D1D5} for some more explanation of the effective string picture of the CFT.} Beyond the cosmic censorship bound, the long string phase ceases to exist and the long string plus short string condensate dominates all the way  to the unitarity bound. This is shown in Figure \ref{fig:MoultingBHs_b}. One finds that the phase boundaries map and the supertube--black hole bound state is identified with the long string--short string condensate. Note that the entropy of the bulk configurations is smaller than that of the CFT phases, which indicates that some of the CFT states are lifted at strong coupling.

Since for a supersymmetric bound state  the energy is exactly the sum of the energy of the constituents, there is no binding energy and the thermodynamic instability does  not imply that a highly rotating BMPV would dynamically expel a supertube, as it cannot lower its energy.\footnote{In  \cite{Bena:2004de,Marolf:2005cx} this system was studied in the probe limit  and indeed this lack of instability showed up as the bound state being marginally bound (no binding energy).}

One can perform a more general analysis than the one in \cite{Bena:2011zw} in the D1-D5 CFT by turning on left and right-moving excitations. This breaks all supersymmetry and makes the system non-extremal. We show below that this system has the same thermodynamic instability on the CFT side as was the case for   \cite{Bena:2011zw}. The lack of supersymmetry suggests that this should map to  an instability in the bulk, where the black hole wants to form a stable bound state with the supertube. In section \ref{s:Minima}, we have found such an instability for supertubes in the non-extremal Cvetic-Youm background. However, in the generic non-extremal black hole charges and anti-charges are excited for all three modes $\bQ_1,\bQ_2,\bQ_3$ and this does not map to the well-understood D1-D5 CFT.\footnote{This is because two of the background charges map to the central charge of the CFT and the third maps to excitations of the CFT. To map to the CFT we need only one kind of charge and anti-charge.} We first need to perform the correct decoupling limit such that only the third charge is excited and we can compare to the D1-D5 CFT.  After taking this decoupling limit we expect the CFT thermodynamic instability to map to the bulk physics by having supertubes co-rotating with the black hole forming a stable bound state  (negative binding energy).  We find that there are stable bound states for large angular momentum. 

\subsubsection{D1-D5 frame}
\label{sss:D1D5}

We discuss how the  Cvetic-Youm in the M-theory frame maps to the D1-D5 frame.  The 11-dimensional solution \bref{eq:11d_Background} can be dimensionally reduced along one of the directions of the first  torus to give an $S^1 \times T^4$ compactification. This system can then, by a series of T and S dualities, be mapped to a frame where the three M2 brane charges become $D1$ wrapping the $S^1$, D5 wrapping the $S^1 \times T^4$ and momentum along the $S^1$. This frame is called the D1-D5 frame. If we take a decoupling limit by zooming into the core of the geometry in the D1-D5 frame, as described in more detail in section \ref{ssec:DecouplingNonExtr}, we get an asymptotically $AdS_3 \times S^3$ region. At the same time the Higgs branch of the D1-D5 system flows in the IR to what is known as the D1-D5 CFT.  This CFT is believed to be marginally deformable to the so called {\em orbifold point}. The supergravity point in the moduli space of the D1-D5 system is not the same as the orbifold point but many calculations done at the orbifold point are seen to match the results at the gravity point exactly \cite{Das:1996wn,Maldacena:1996ix,David:1998ev,Lunin:2001dt,Avery:2009tu,Avery:2009xr} and some results get corrections but the qualitative behavior is the same  \cite{Giusto:2004id, Bena:2011zw}. 
The remarkable ability of calculations done at orbifold point to capture supergravity physics and its considerably simplicity make it an invaluable tool.  According to the AdS/CFT paradigm the asymptotic $AdS_3 \times S^3$ and the D1-D5 CFT are dual to each other. We will study the D1-D5 CFT at the orbifold point. Technical details of the D1-D5 can be found in appendix \ref{section:D1D5CFT}.

We will use the crude description of the D1-D5 CFT at the orbifold point in terms of `effective strings'. These effective strings have a total winding number the product of the number of D1 and D5 branes and runs along $S^1$. The total winding can be distributed into many singly or multiply wound effective strings. The excitations of this effective string run up and down $S^1$ and are called left movers and right movers.

\subsection{Bound  states in CFT}

In this section, we show there exists a \emph{thermodynamical} instability at the orbifold point in the D1-D5 CFT. Such a thermodynamic instability was also found in the supersymmetric limit in \cite{Bena:2011zw}. Note that since are no dynamics at the oribifold point and  a thermodynamic instability is the only type of instability we can find. We hope to  show a dynamic instability using recent techniques of moving off the orbifold point discussed in \cite{Avery:2010er,Avery:2010hs,Avery:2010vk} in the future. 

We discuss the various phases in terms of the intensive charges of the corresponding CFT state:
\begin{gather}
n_{3L} = \f{\bN_{3L}}{\bN}\,, \qquad n_{3R} = \f{\bN_{3R}}{\bN}\,, \\
j_{L} = \f{ J_{L}}{\bN}\,, \qquad j_{R} = \f{ J_{R}}{\bN}\,, \label{eq:IntensiveQuantities}
\end{gather}
where $6 \bN$ is the central charge of the D1-D5 orbifold CFT, $\bN_{3L},\bN_{3R}$ are the (normalized) left- and right-moving conformal dimensions of the state and $J_L,J_R$ its R-charges. See appendix \ref{section:D1D5CFT} for more information on the D1-D5 CFT.\footnote{Note that in appendix \ref{section:D1D5CFT}, we choose standard notation for the CFT charges. In particular, we denote $N_{pL},N_{pR}$ instead of $N_{3L},N_{3R}$.}

\medskip

\begin{figure}
\centering
\subfigure[BMPV phase]{
\includegraphics[width=.35\textwidth]{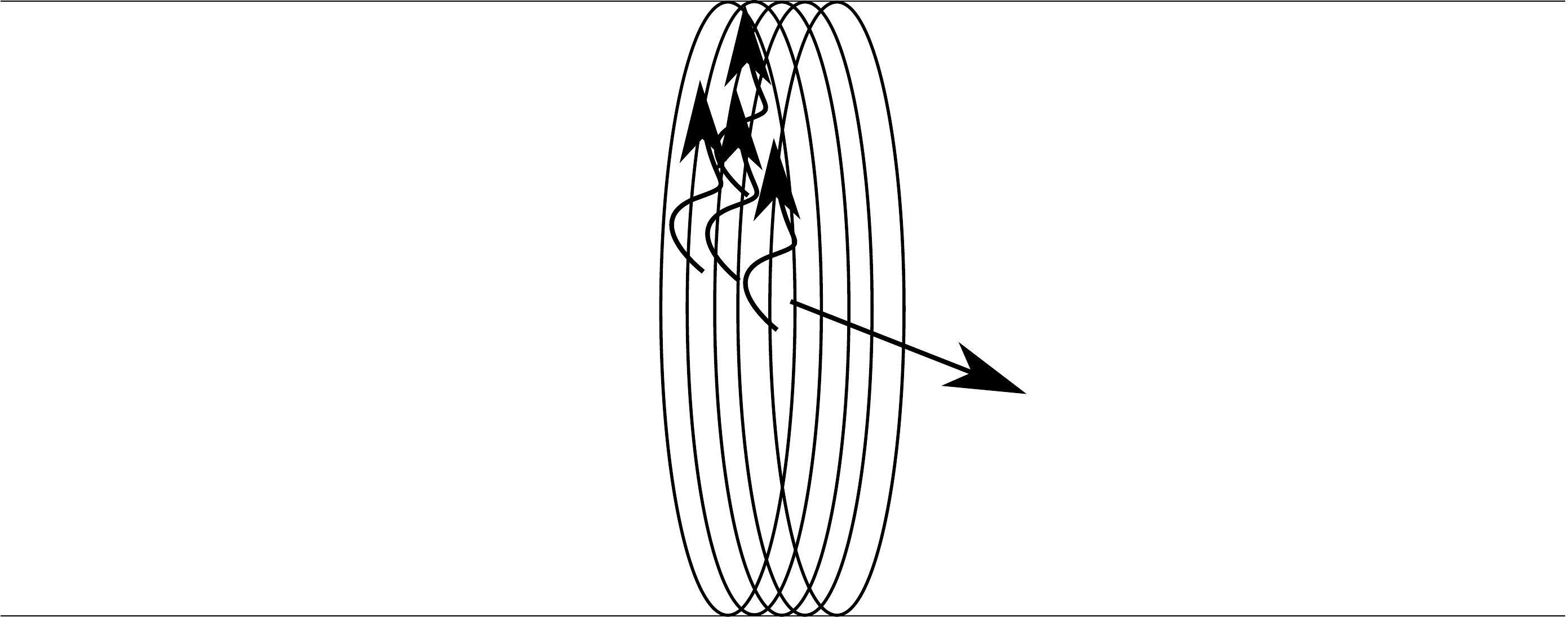}
\label{fig:3phases_a}
}
\hspace{.05\textwidth}
\subfigure[Enigmatic phase]{
\includegraphics[width=.35\textwidth]{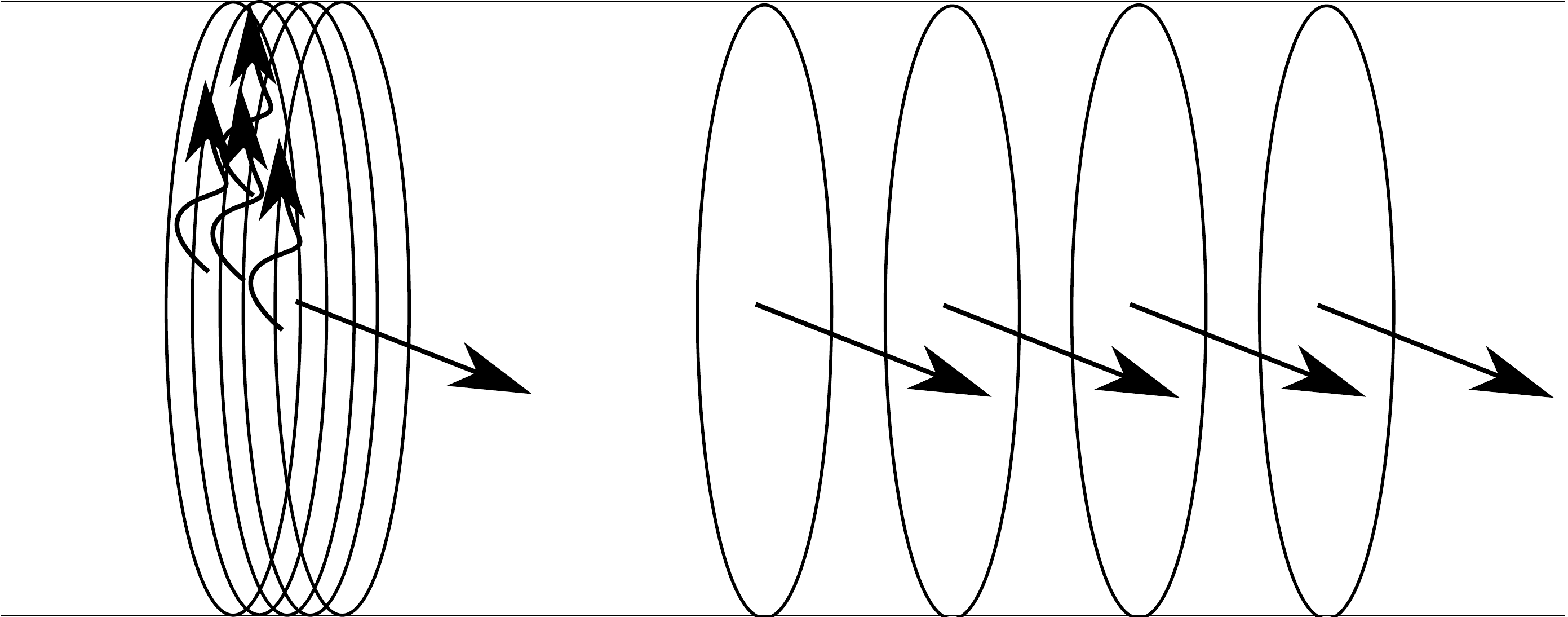}
\label{fig:3phases_b}
}
\caption{\sl The BMPV and the enigmatic phases at the orbifold point of the D1-D5 CFT.\label{fig:3phases}}
 \end{figure}

Let us first review the instability found in \cite{Bena:2011zw}. The BMPV black hole is dual to states in the CFT that, to leading order,  are described by left-moving excitations on one long string of winding $\bN$. This phase is shown in Figure \ref{fig:3phases_a} and has an entropy
\be
S_{BMPV}=2 \pi \bN \sqrt{ n_{3L} - j_L^2}\,.
\ee

However, there is another phase possible when $\bN l$ windings condense into singly twisted sectors with no excitations on them. The left moving angular momentum of the short string condensate is taken to be aligned with the total angular momentum. This part carries no entropy and all the entropy is carried by the remaining long string
\be
S_l=2\pi \bN \sqrt{(1-l) n_{3L} - (j_L-\f{l}{2})^2}\,,
\ee
where each short string carries away one unit of winding and half a unit of angular momentum in the left and right sector.\footnote{In \cite{Bena:2011zw} the chemical potential for $j_R$ was zero so the two phases with different $j_R$ had the same energy.} This entropy is maximized for
\be
l=2(j_L- n_{3L})\,,
\ee
and its maximal value is
\be
S_{enigma}=2 \pi \bN \sqrt{n_{3L} (n_{3L}+1-2 j_L)} \,.
\ee
This new phase is shown in Figure  \ref{fig:3phases_b}.  We follow \cite{Bena:2011zw} and name this the `enigmatic' phase, as it is related to the entropy enigma of \cite{Denef:2007vg,deBoer:2008fk}, where certain two-center
BPS black hole configurations can have larger entropy than a single-center solution with the
same asymptotic charges.

When $j_L > n_3$, the entropy at the orbifold point for the enigmatic phase is larger than the one for the BMPV phase and there is a thermodynamical instability, favouring the enigmatic phase.

\medskip

Now we repeat this analysis for non-extremal charges in the CFT. This is done by having both left and right movers turned on in the CFT. At leading order such a black hole is dual to states in the CFT which have left and right-moving excitations on the long string of winding $\bN$. At the orbifold point the left and right movers do not interact so the entropy is additive
\be
S_{BH}= 2 \pi \bN   \sqrt{n_{3L} -  j_L^2}\,\, + \,\, 2 \pi \bN  \sqrt{n_{3R} -  j_R^2}\,.
\ee
We can look at states where some winding is taken out of the long string and we have in addition to a long string carrying all the excitations a condensate of short strings. Each short string caries $ \pm \h$ units of angular momentum in the left and right sectors. If the total length of the short strings is $\bN l$ then the entropy is
\be
S_l=2 \pi \bN \sqrt{(1-l) n_{3L} -  (j_L +j_L^{\rm short})^2}+ 2 \pi \bN \sqrt{(1-l) n_{3R} -  (j_R +j_R^{\rm short})^2}\,,
\ee
where $j_L^{\rm short} = \pm \f{l}{2}$ and $j_R^{\rm short} = \pm \f l 2$ (independent signs) to account for the possible relative orientations of the short string condensate angular momenta with the total angular momenta.  When both signs are equal, the short string condensate angular momentum is (anti-)aligned with the total angular momentum. When they have opposite signs then the short string condensate has angular momentum orthogonal to the total angular momentum.
 
We can maximize this entropy with respect to $l$ but the answer is complicated and not particularly illuminating.  For illustrative purposes and comparison with the bulk physics we concentrate on states with charges
\be
n_{3L}= n_{3R}=n_3, \qquad j_L=-j_R=j\,,
\ee
which corresponds in the bulk to a black hole with no net third charge and rotating in $\psi$ plane only.\footnote{The relative sign of $j_R,j_L$ determine whether this is the $\psi$- or the $\phi$-plane.} We call such a CFT state a black hole state. It is shown in Figure \ref{fig:non-extremalBHCFT} and its entropy is
\be
S_{BH}=4 \pi \bN  \sqrt{n_3 - j^2}\,.
\ee
We discuss two possibilities for the entropy of the enigmatic state, depending on the orientation of the short strings' angular momentum $j^{\rm short}_L=\mp j^{\rm short}_R$.

\subsubsection{Angular momentum of short strings and background (anti)aligned}

\begin{figure}
\centering
\subfigure[Non-Extremal Black Hole]{
\includegraphics[width=.35\textwidth]{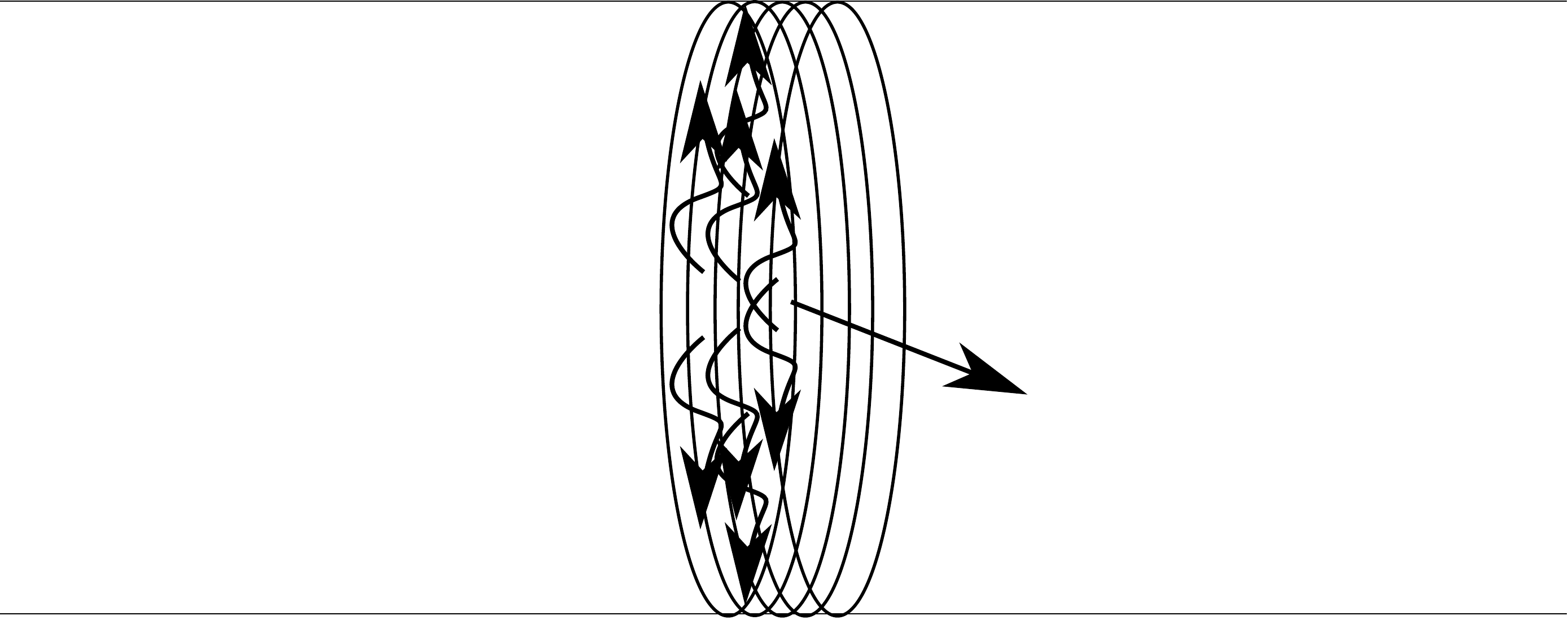}
\label{fig:non-extremalBHCFT}
}
\hspace{.05\textwidth}
\subfigure[Non-Extremal Enigmatic Phase]{
\includegraphics[width=.35\textwidth]{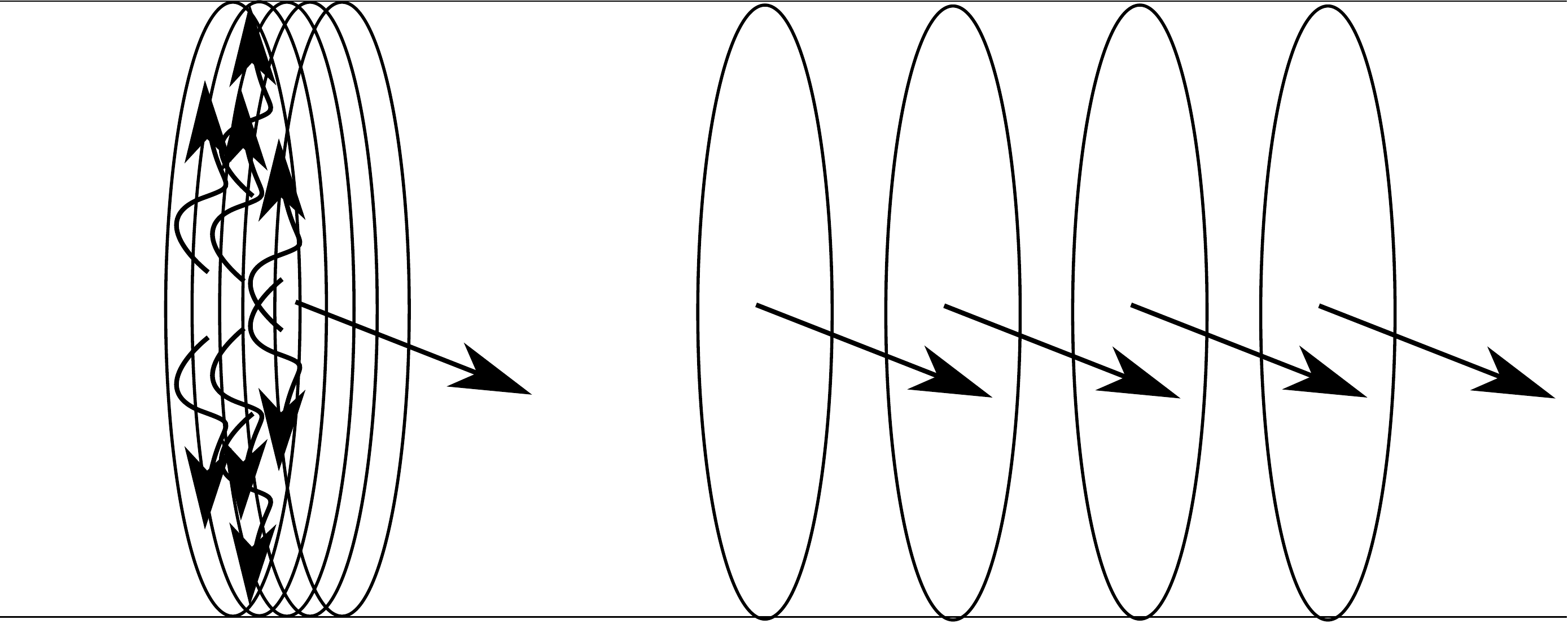}
\label{fig:non-extremalEnigmaticCFT}
}
\caption{\sl The non-extremal black hole with no net third charge and the corresponding enigmatic phase at the orbifold point of the D1-D5 CFT.\label{fig:phasesNon-Extremal}}
 \end{figure}

If the short strings have their angular momentum (anti) aligned with the total angular momentum, $j^{\rm short}_L=-j^{\rm short}_R = l/2$, the entropy of the new phase is
\be
S_l=4 \pi \bN \sqrt{(1-l) n_3 - (j -\f{l}{2})^2}\,,
\ee
which is maximized for 
\be
l=2(j- n_3)\,,
\ee
and is
\be
S_{enigma}=4 \pi \bN \sqrt{n_3 (n_3+1 - 2 j)}\,.
\ee
This phase is dual to the bulk configuration depicted in Figure \ref{fig:tube-same-plane}.
These states are shown in Figure \ref{fig:non-extremalEnigmaticCFT}.
The bound for existence and dominance of the new phase ($l>0$ or $j>n_3$) is the same as that for the supersymmetric case studied in \cite{Bena:2011zw} due to the charges in the left and right sector being equal. This solution only exists for $j>0$.

\begin{figure}[tbh]
 \centering
   \includegraphics[width=.6\textwidth]{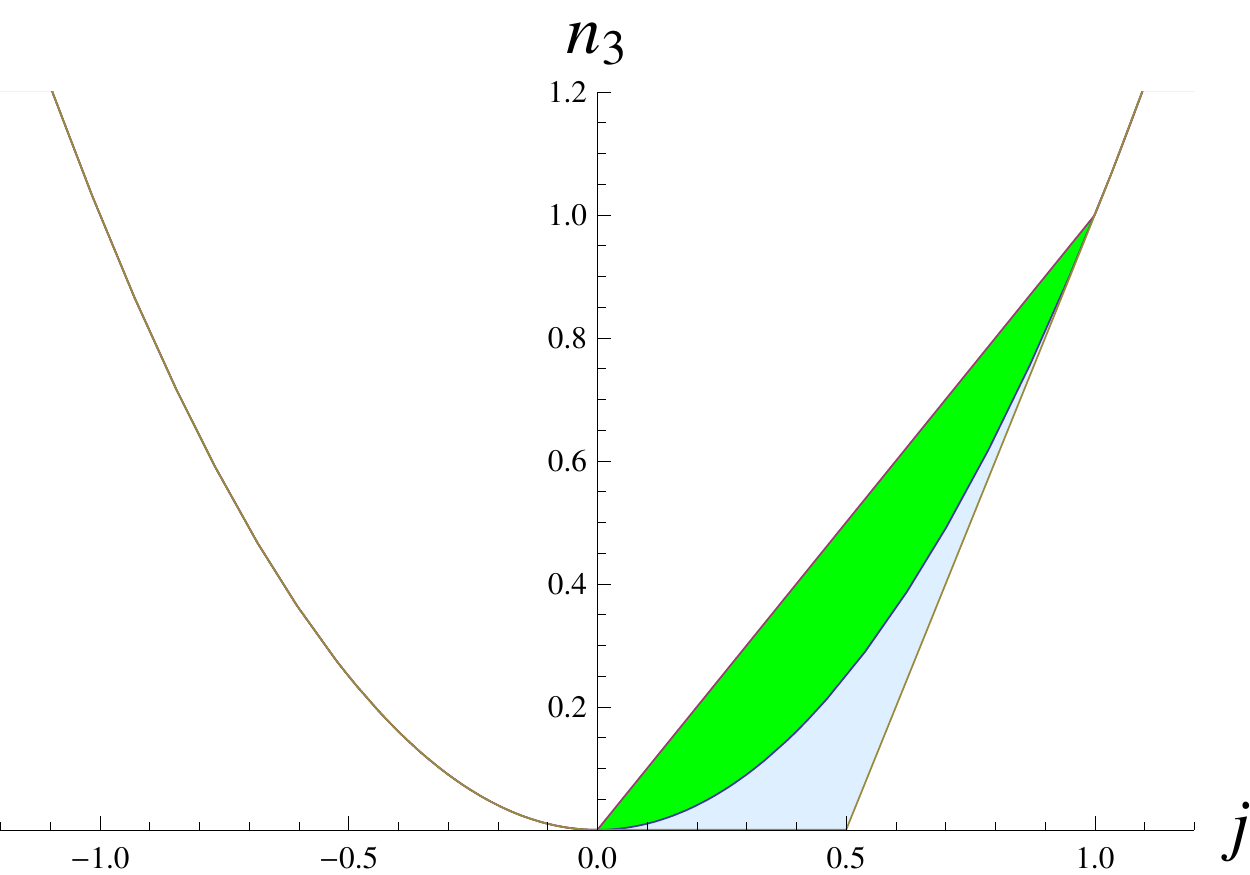}
   \caption{The black hole exists above the cosmic censorship bound, $n_3 \ge j^2$. The green region ($j>n_3$ and $n_3>j^2$) is where the enigmatic phases exists and dominates . This is consistent with the intuition that the black hole wants to spit out angular momentum. On the CFT side we are working outside the probe limit so we also get the blue region ($n_3>0, n_3<j^2$ and $n_3>2 j-1$) where the long string sector does not exist but the enigmatic phase still exists and dominates on other phases.
  \label{fig:2-charge-existence-region}}
\end{figure}

\subsubsection{Angular momentum of short strings orthogonal to total angular momentum}

If the short strings have their angular momentum orthogonal to the total angular momentum, $j^{\rm short}_L=+j^{\rm short}_R = l/2$, the entropy of the new phase is
\be
S_l=2 \pi \bN \sqrt{(1-l) n_3 - (j- \f{l}{2})^2}+ 2 \pi \bN\sqrt{(1-l) n_3 - (j+ \f{l}{2})^2}\,.
\ee
On taking the derivative with respect to $l$ we get two solutions
\be
l_\pm= 1- n_3 \pm  \sqrt{(1+n_3)^2-4j^2}\,.
\ee
These roots are always defined, as the unitarity bound in the CFT requires $n_3 >  2|j|-1$ .

The entropy for these roots is
\be
S_\pm = \sqrt{2} \pi \bN ( \sqrt{ -(n_3+1  - 2j) l_\pm} + \sqrt{-(n_3+1+2j) l_\pm} ) \,.
\ee
Since $l$, the number of short strings, is positive, the entropy is imaginary. We conclude there is no enigmatic phase for this choice of angular momentum. This phase is dual to the bulk configuration depicted in Figure \ref{fig:tube-orthogonal-plane}.

\subsection{Decoupling limit of the non-extremal black hole}
\label{ssec:DecouplingNonExtr}

We wish  to interpret the CFT results in the dual geometry. Therefore we study the D1-D5 decoupling limit of the non-extremal black hole metric \bref{eq:11d_Background} in the D1-D5 frame. It is a product of $T^4$ with  twisted fibration of $S^3$ over a BTZ black hole \cite{Cvetic:1998xh}. This allows us to map the various stable, marginal and metastable bound states we find in the bulk to phases in the dual D1-D5 CFT.

This decoupling limit is obtained by assuming $\bQ_1,\bQ_2 \gg \bQ_3,m,a_1^2,a_2^2$, such that we can study non-extremal excitations of $\bQ_3$ in the $\bQ_1,\bQ_2$--background, and focussing on the region $r^2 \ll \bQ_1,\bQ_2$.  When $m=0$, the system is dual to supersymmetric (only left- or right-moving) excitations of the D1-D5 CFT with central charge $6 \bN_1 \bN_2$, where $\bN_I$ denote the integer charges of the background. We want any non-extremality  ($m>0$) to excite only the third charge so that this maps to both left-and right-moving excitations in the same CFT dual. These conditions are captured by
\be
{\bQ_3}^2 \ll \bQ_1 \bQ_2, \qquad m_3^2 \ll m_1 m_2.
\ee
where $m_I$ are the mass of M2 branes wrapped on the $I^{th}$ torus defined in \bref{massCharges}.
We implement the second condition by defining a small parameter $\epsilon$ and taking
\be
m_1=m_2=1, \, m_3=\epsilon\,,
\ee
and expanding the Hamiltonian to first order in $\epsilon$. The first condition is then automatically satisfied\footnote{In addition to the above conditions we require the usual AdS/CFT conditions:  the AdS radius is large in string units and very large in Planck units in the D1-D5 frame. If the $T^6$ radii are given by $R_x, R_z, R_5, R_6, R_7, R_8$ with reduction to type IIA on direction $x$ then these conditions are respectively $ \f{\bN l_{11}^6}{R_z^2 R_5 R_6 R_7 R_8} \gg 1$, $\bN \to \infty$ where $l_{11}$ is the 11-dimensional Planck length.}.
With this convention the first two harmonic functions in the core region are (in terms of the integer charges)
\be
H_1 = \f{\bN_1}{\epsilon f}, \qquad H_2 =  \f{\bN_2}{\epsilon f}\,.
\ee

In the D1-D5 frame, the metric then describes small $\bQ_3$ and anti-$\bQ_3$ excitations on a warped $BTZ \times S^3 \times T^4$ geometry. The third charge and anti-charge map to left and right-moving excitations of the D1-D5 CFT dual  and the angular momenta to the R-charges of this CFT. The CFT charges can be read off from the gravity quantities as 
\begin{gather}
\bN_{3L} = \f{1}{4} m e^{2 \delta_3}\,, \qquad \bN_{3R} = \f{1}{4} m e^{-2 \delta_3}\,, \nn
J_L = -\h {\sqrt{\bN_1 \bN_2}} (a_1 -a_2) e^{\delta_3}\,, \qquad J_R = -\h {\sqrt{\bN_1 \bN_2}} (a_1 +a_2) e^{-\delta_3}\,.
\end{gather}
In principle, we can invert these relations and write the Hamiltonian explicitly in terms of the CFT variables. However, this form is not in general particularly illuminating except in a some special cases which we give below. 

Since we want to compare to the orbifold point in the dual CFT which only sees the combination $\bN_1 \bN_2$ we make the simplifying assumption
\be
\bN_1 = \bN_2 = \sqrt{\bN}, \quad \pN_1 =\pN_2 = \sqrt{N_{\rm tube}}\,.
\ee
which does not change the physics qualitatively. 

\medskip

To first order in $\epsilon$, the Hamiltonian is
\bea
\calh - 2 \sqrt{N_{\rm tube}}  &=\frac{\epsilon}{\sqrt{\bN}} f \bigg(   &\frac{\sqrt{H_m}( \sqrt{N_{\rm tube}}  - s_3 \beta +c_3 \alpha)^2}{\sqrt{ H_3 g_{\a\a}^{(4)}}-\sqrt{H_m} (s_3 \beta- H_m^{-1} c_3 \alpha)}\nn
&&-(\sqrt{N_{\rm tube}}  -s_3 \beta)- H_m(\sqrt{N_{\rm tube}} + c_3 \a)
+\sqrt{H_m H_3 g_{\a \a}^{(4)}} \bigg)\,.  \nonumber
\eea
where
\begin{gather}
H_m = 1- \f{4\sqrt{\bN_{3L} \bN_{3R}}}{f}\,, \qquad H_3 = 1+ \f{ ( \sqrt{\bN_{3L}} - \sqrt{\bN_{3R}})^2}{f}\,, \nn
\alpha=a_1 b_1 \cos^2 \theta + a_2 b_2 \sin^2 \theta\,, \qquad \beta=a_2 b_1 \cos^2 \theta + a_1 b_2 \sin^2 \theta\,,\nn
f=r^2 +a_1^2 \sin^2 \theta + a_2^2 \cos^2 \theta\,,\qquad g_{\a\a}^{(4)} = f(b_1^2 \cos^2 \theta + b_2^2 \sin^2 \theta) + H_m^{-1} \alpha^2 - \beta^2 \,. 
\end{gather}
It is not hard to see that  the shifted and scaled Hamiltonian
\be
\mathscr H = \f{\calh - 2 \sqrt{N_{\rm tube}}}{\sqrt{\bN} \epsilon} \label{eq:Ham_Intensive}
\ee
is a function only of the intensive quantities defined in \bref{eq:IntensiveQuantities} and of 
\begin{gather}
{\hat r}^2\equiv\frac{{r^2}}{{\bN}}, \qquad \pn \equiv   \sqrt{\f{N_{\rm tube}}{\bN}}\,.
\end{gather}
This follows from the scaling property discussed in the appendix, eq.\ \eqref{eq:Scaling_2}. Thus the intensive Hamiltonian $\mathscr H$ has a natural interpretation in the CFT.

\medskip

To discuss the stability properties, we specialize to the case where the net third charge is zero ($s_3=0$) and the angular momentum is only in the $\psi$ plane ($a_2=0$). Then we have
\be
n_{3L} = n_{3R} \equiv n_3, \qquad j_L = -j_R\equiv j \,,
\ee
This restriction keeps the rich physics, but makes an analytical treatment possible. 

We study the two main possibilities for the orientation of the tube angular momentum:  first we let the tube rotate in the same plane as that of the black hole, then we give it rotation only in the orthogonal plane. Both have metastable bound states, but only when the tube and black hole rotate in the same plane, the bound state can become stable. Note that in contrast to previous section, we keep the background fixed and change the embedding of tube to describe different relative orientations of angular momenta.

\subsubsection{Supertube rotation in same plane as that of black hole}\label{sss:Decoupling_SamePlane}

To make the tube rotate in the same plane as that of the black hole we take $\theta=0$, $b_1=1$ and $b_2=0$. There are two possibilities, one where the tube co-rotates with the black hole ($j>0$) as shown in Figure \ref{fig:tube-same-plane_a} and the  other where the tube rotates against the black hole ($j<0$) as shown in Figure \ref{fig:tube-same-plane_b}. 

The Hamiltonian is
\be
 \mathscr H =
 (\hrho+2 j )(\hrho -\pn )^2+4 n_3(\hrho -\pn) \,,\label{eq:Hamiltonian_Decoupling_SamePlane}
\ee
where we have used the shifted radial coordinate
\be
\hat {\rho}^2 = {\hat r}^2-4 ( n_3- j^2) \,,
\ee
which measures the distance from the horizon. The local maxima and minima of the potential \eqref{eq:Hamiltonian_Decoupling_SamePlane} occur at
\be
\hrho_{-} = \f{1}{3} \left(2 ( \pn-j) - \sqrt{(2j+\pn)^2-12 n_3}\right), \qquad \hrho_{+} = \left(\f{1}{3} (2 ( \pn-j) + \sqrt{(2j+\pn)^2-12 n_3}\right)\,,\label{eq:RhoPlusMin}
\ee
respectively and a bound states exist when the discriminant $D^2$ is positive:
\be
D^2 \equiv (2 j+ \pn)^2 -12 n_3 \ge 0\,.
\ee
The `binding energy' is given by the difference between the potential at the local minimum and the horizon
\be
\mathscr H |_ {\hrho = \hrho_+} - \mathscr H| _{\hrho =0}=-\f{2}{27} (D-2(j-\pn))^2(D+(j-\pn))  \,,
\ee
which should be negative for stability. Hence the condition for a stable bound state is
\be
D> \pn-j\,.
\ee

\begin{figure}[tbh]
\centering
\epsfxsize=7cm \epsfbox{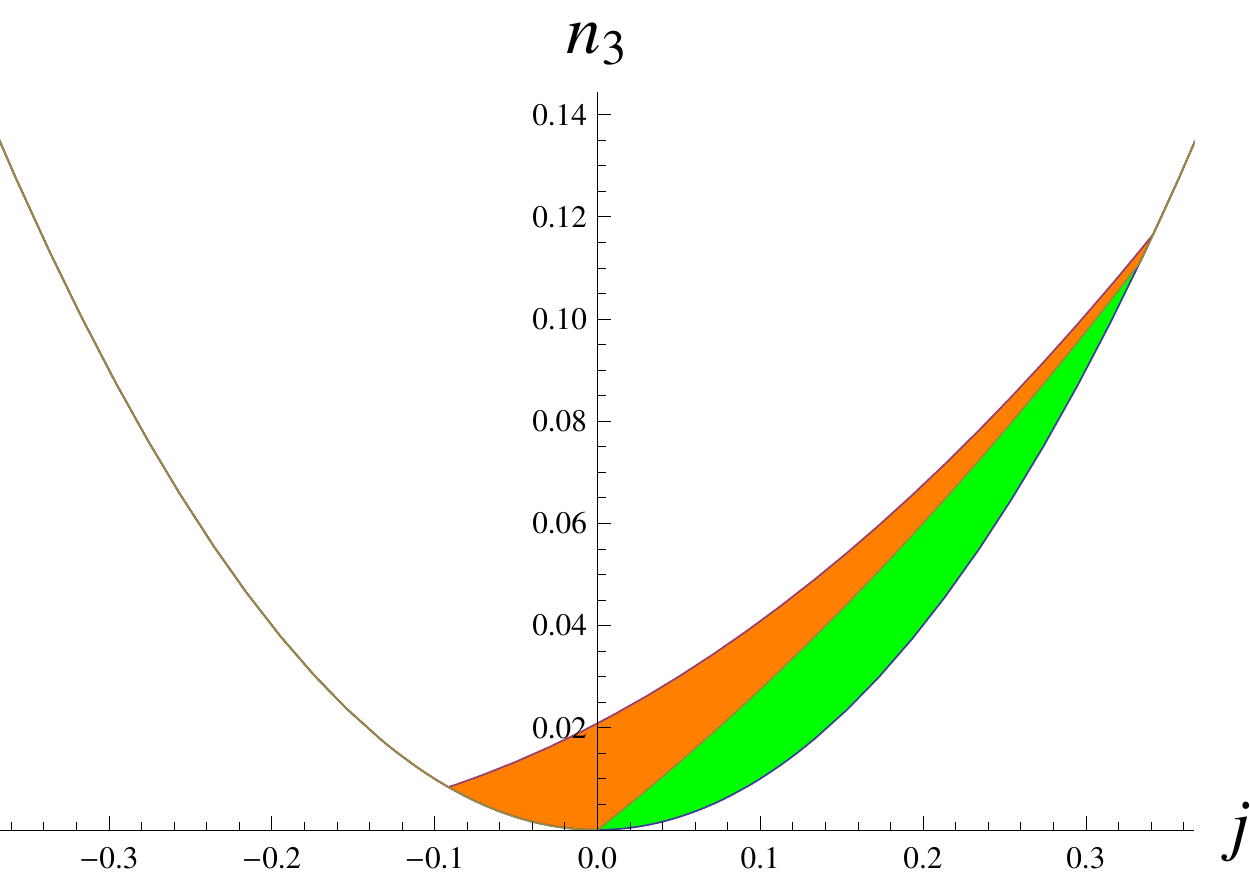}
\caption{\sl The phase diagram in the decoupling limit for black hole--supertube bound states with angular momentum in the same plane,  for $\pn=\h$. The background black hole exists (has no naked CTC's) above the cosmic censorship parabola, $n_3 \ge j^2$. Metastable bound states exist in the orange region for $n_3 < \tfrac 13 (j + \pn/2)^2$, stable bound states exist in the green region, with $n_3 <\tfrac 14 (j^2 + 2\pn j)$.
  \label{fig:2-charge-co-planar-existence-region}}
\end{figure}

From these conditions, we find three distinct regions in the phase diagram for the background charge $n_3$ vs.\ the background angular momentum $j$: the region with stable bound states, the region with metastable bound states and the region with no bound states. The bounds for these regions depend on the charge $\pn$ of the probe: for smaller $\pn$, the size of the different regions shrink. We gather these bounds and show these regions in Figure \ref{fig:2-charge-co-planar-existence-region} for a representative value $\pn=0.5$. Although this value is too large for a probe the qualitative features do not depend on the value of $\pn$. 

Note the difference when the tube and the background anti-rotate $(j<0)$ or co-rotate $(j>0)$.  Only when $j>0$, one observes stable bound states. When $j<0$, the angular momentum interaction is attractive and there are no stable bound states. This is consistent with the intuition that the black hole wants to expel angular momentum. It is instructive to observe that there are still metastable bound states for both $j>0$ and $j<0$.

\subsubsection{Supertube and black hole rotating in orthogonal planes}

To make the supertube rotate in a plane orthogonal to that of the rotation of the black hole we take $\theta= \f{\pi}{2}$, $b_1=0$ and $b_2=1$. This configuration is shown in Figure \ref{fig:tube-orthogonal-plane}. 

Then the Hamiltonian is
\be
\mathscr H =  \f{\rho \sqrt{\hrho^2 + 4n_3} }{\sqrt{ \hrho^2 + 4(n_3-j^2)}}  ( \hrho^2 + 4(n_3-j^2)+\pn^2 ) - 2 (\hrho^2+2 n_3) \pn  \,,
\ee
again in terms of the shifted radial coordinate
\be
\hat {\rho} = {\hat r}^2-4 ( n_3- j^2) \,,
\ee
which measures the distance from the horizon. 

This expression is not amenable to a full analytic study. However, on the cosmic censorhip parabola ($n_3 = j^2$), the Hamiltonian simplifies and one finds bound states for
\be
|j|\leq j_{\rm max} \equiv  \frac{2 + \sqrt 2}{4} \pn\,.
\ee
Raising  $n_3 >j^2$ adds more mass to the system (but not more overall charge) and will lift the energy of the bound state. 

\begin{figure}[tbh]
\begin{center}
\includegraphics[width=7cm]{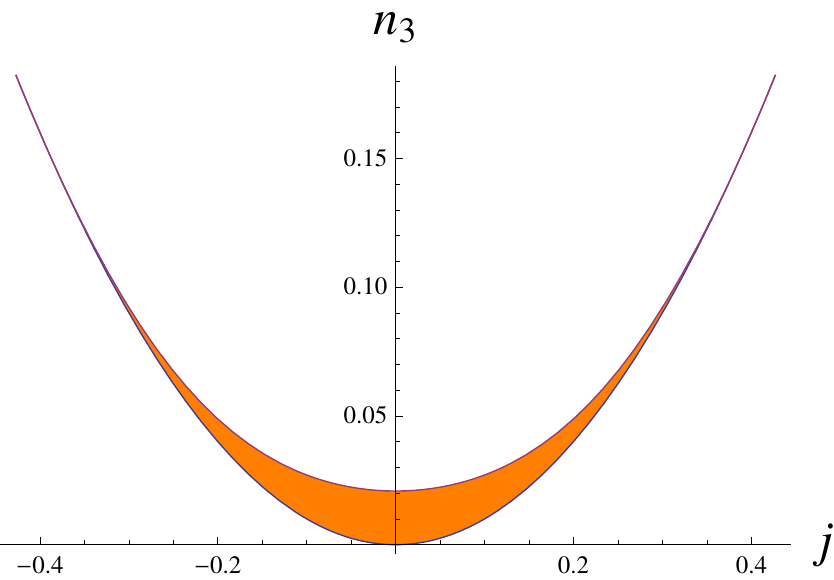}
\caption{\sl The phase diagram in the decoupling limit for black hole--supertube bound states with angular momentum in \emph{orthogonal} planes,  for $\pn=\h$.  The background black hole exists above the cosmic censorship parabola,  $n_3\geq j^2$. The orange region is where supertubes can form a metastable bound state with the black hole. All bound states have $j\leq j_{\rm max} \approx 0.4268$.  There are no stable bound  states.
\label{fig:2-charge-orthogonal-existence-region}}
\end{center}
\end{figure}

For generic values of  $n_3$, we do not have analytic control so we study the $n_3$ vs. $j$ phase diagram numerically. For $n_3>j$ there is indeed a small region where {\em metastable} bound states exist. The phase diagram is plotted for an illustrative example in Figure \ref{fig:2-charge-orthogonal-existence-region}. Note the symmetry $j \leftrightarrow -j$. This is consistent with the intuition that since the black hole rotation and the supertube rotation are in orthogonal planes the sign of the black hole angular momentum does not matter. Note also that there are no stable bound states

\subsection{Comparison of results from the bulk and the boundary}

\begin{figure}[htbp] 
   \centering
\subfigure[Bulk: Rotation in same plane \label{fig:BulkRotationInSamePlane}]{
\includegraphics[width=7cm]{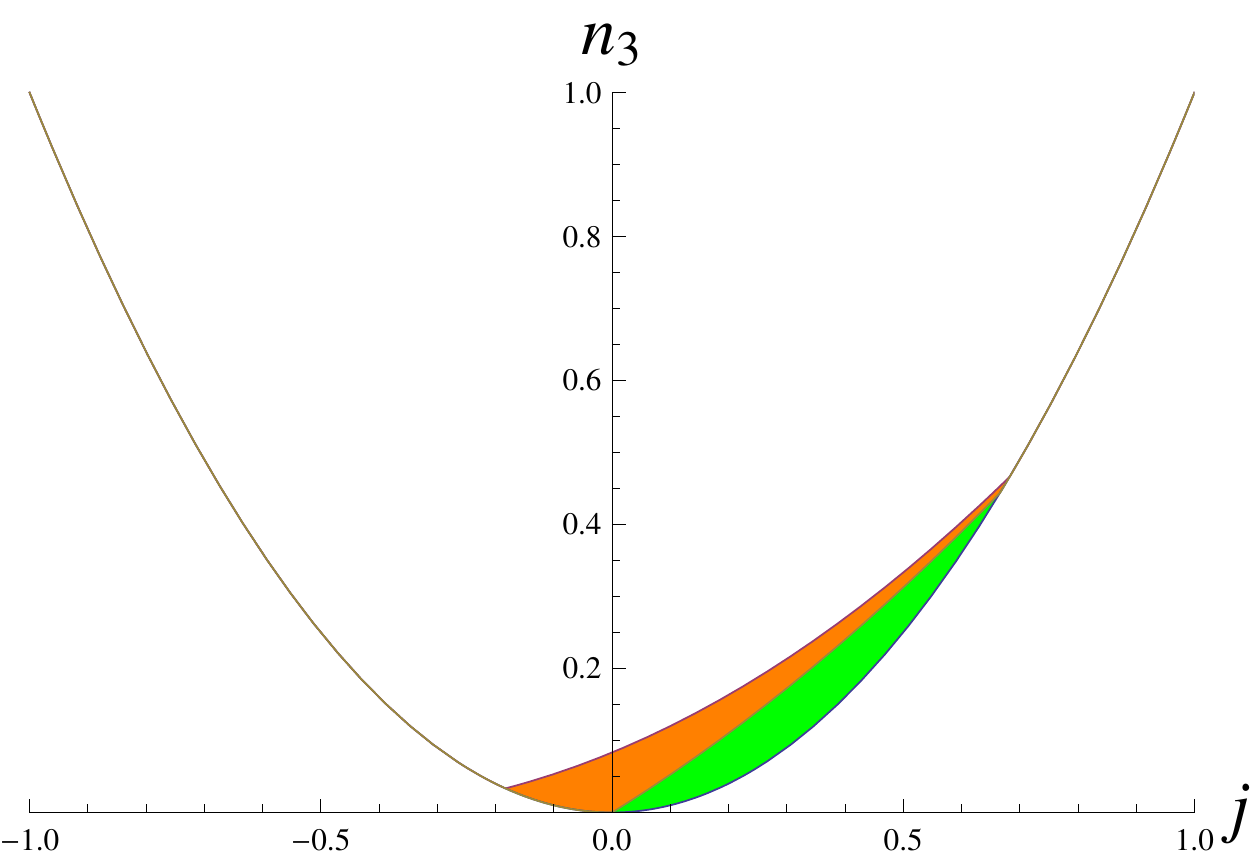}
}
\subfigure[CFT: Rotation in same plane.\label{fig:CFTRotationInSamePlaneb}]{
\includegraphics[width=7cm]{Images/2-charge-existence-region-CFT}
} \\
\subfigure[Bulk: Rotation in orthogonal plane.\label{fig:BulkRotationInOrthogonalPlaneb}]{
\includegraphics[width=7cm]{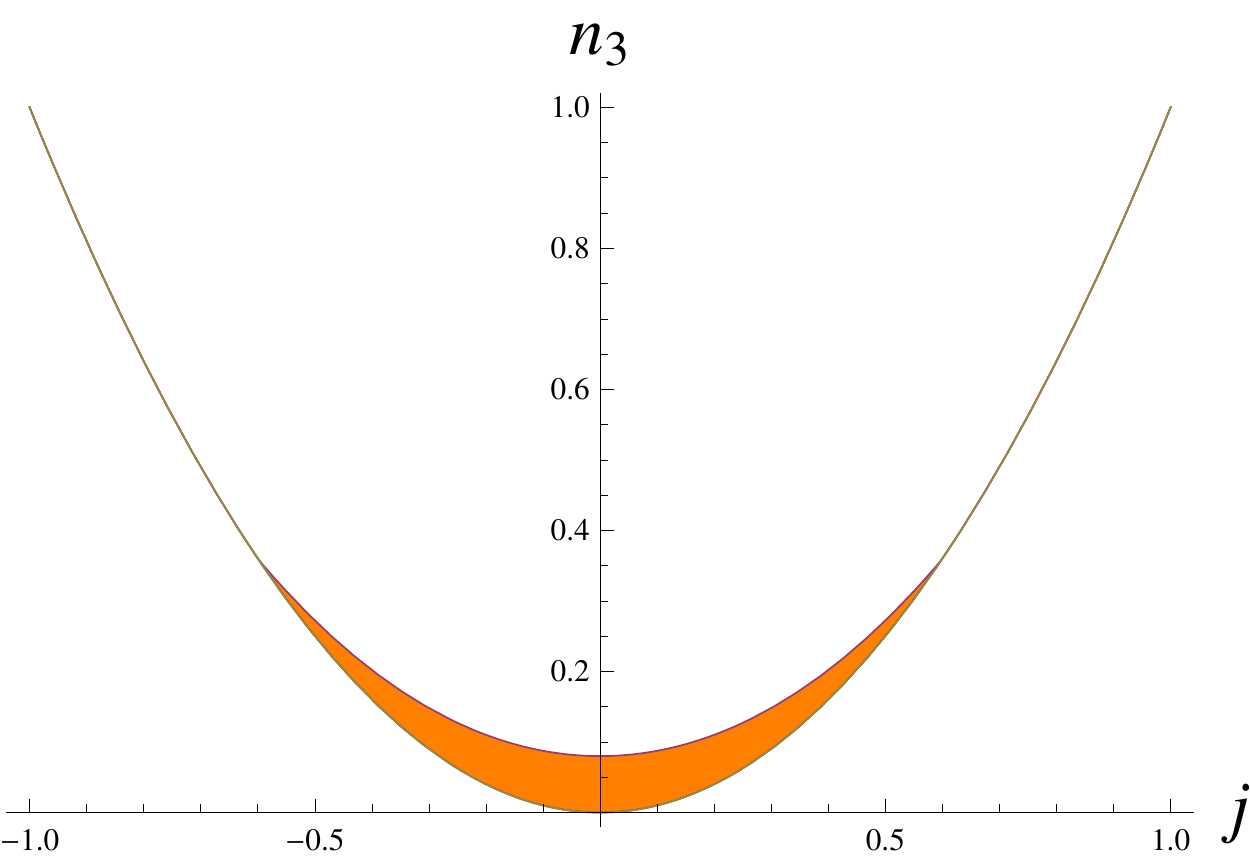}
}
\subfigure[CFT: Rotation in orthogonal plane.\label{fig:CFTRotationInOrthogonalPlane}]{
\includegraphics[width=7cm]{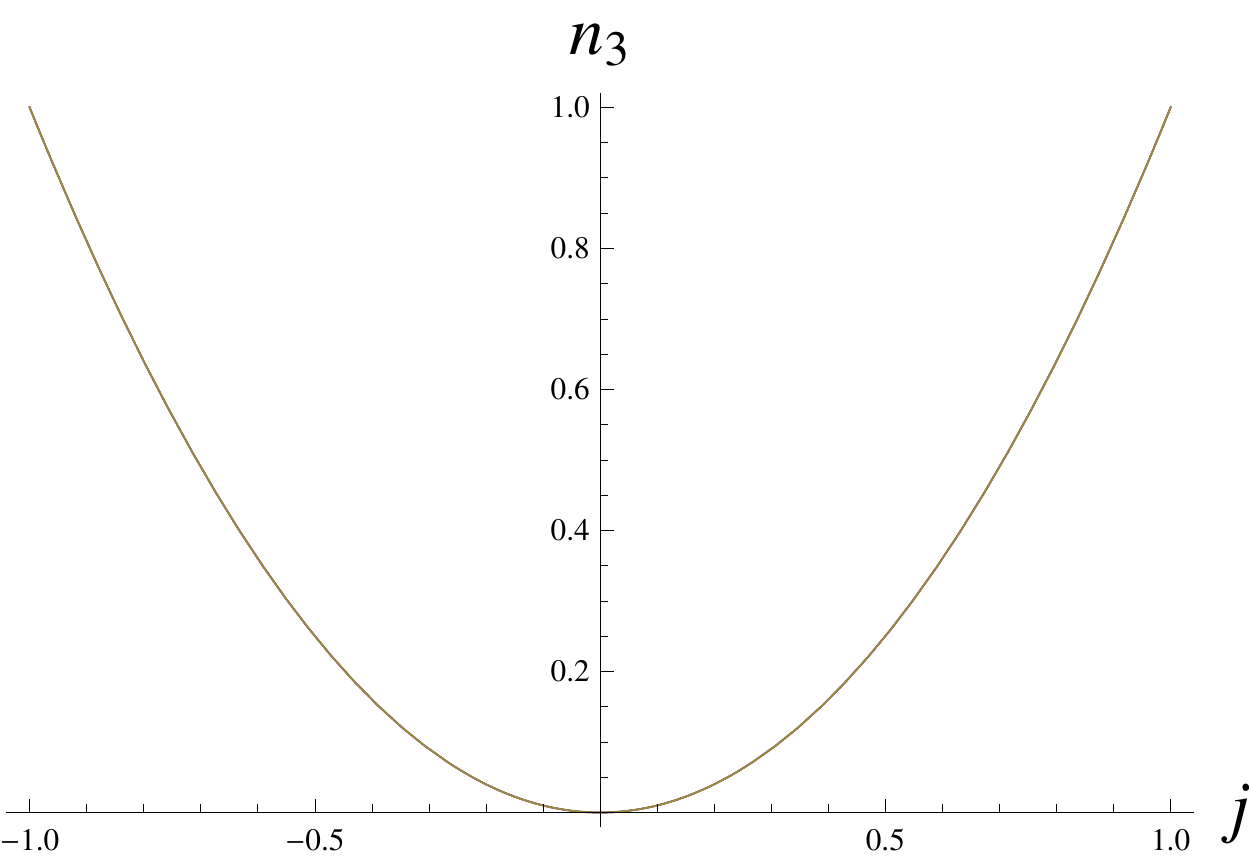}
}
   \caption{\sl The phase diagrams from bulk and boundary analysis. From the bulk we find stable (green region) states for supertubes rotating along the black hole and metastable states (orange region) for supertubes rotating along and against the black hole (Figure (a)). Furthermore, for supertubes rotating in a plane orthogonal to the rotation of the black holes we find only metastable states (orange region in Figure (c)). From the boundary we find that the new `enigmatic' phase has a larger entropy than the black hole phase stable states for rotation along the black hole (green region in Figure (b)), but no metastable states for orthogonal rotation (Figure (d)). }
   \label{fig:comparison}
\end{figure}

The CFT analysis was performed at the orbifold point which has a different regime of validity than the supergravity description. Thus it is instructive to  compare the results. The CFT and bulk phase diagrams are plotted in Figure \ref{fig:comparison}. The bulk analysis was done in the probe limit  and therefore the phase diagrams are limited to being inside the cosmic censorship bound  where the black hole exists ($n_3 \geq j^2)$.  We take $\pn=1$ for the supertube charge in Figure \ref{fig:comparison}, even though strictly speaking it falls outside the probe limit, to be able to compare to the CFT where the short string condensate has charges of the same order as the long string.\footnote{The qualitative features of the bulk phase diagrams are independent of $\pn$ but the existence regions of metastable and stable states shrinks with $\pn$; for a more quantitative treatment of the bound states and exact comparison with the CFT, one would need a fully back-reacted analysis.}

We see from the bulk that when the supertube and the black hole have rotation in the same plane, there are stable bound states when the rotations are aligned $(j>0)$. They exist for small $n_3$ and large angular momentum shown by the green region in Figure \ref{fig:BulkRotationInSamePlane}. From the CFT analysis we see there is a similar region, where the two-center `enigmatic' phase is entropically dominant for $j>n_3$, shown by the green region in Figure \ref{fig:CFTRotationInSamePlaneb}. The region of stability of the supertube--black hole stable bound states in the supergravity description falls well within the region where the enigmatic phase is thermodynamically stable in the CFT. We take this as evidence that both phases should be identified and hence a thermodynamical instability in the CFT maps to a dynamical instability in the bulk.

For small $n_3$ there is another small region where metastable states exist in the bulk, shown by the orange region in Figure \ref{fig:BulkRotationInSamePlane}. This region mainly has the angular momenta of the tube and the black hole aligned ($j>0$), but also for anti-aligned angular momenta $j<0$ we find a small set of metastable states. These metastable bound states do not have an analogue in the orbifold point of the CFT. It would be interesting to go beyond the orbifold point to see what the CFT interpretation of these metastable states is.

If we choose orthogonal rotational planes for the black hole and the supertube, there is no new enigmatic phase in the CFT (Figure \ref{fig:CFTRotationInOrthogonalPlane}). However, in the supergravity phase diagram of Figure \ref{fig:BulkRotationInOrthogonalPlaneb}, there is a small region, close to the cosmic censorhip bound, where the supertube can form a metastable bound state with the black hole. Again, it would be interesting to go beyond the CFT orbifold point to understand the CFT nature of this metastable phase.

\section{Discussion and future directions}
\label{s:Discussion}

In this article, we have calculated the potential of supertubes in the non-extremal Cvetic-Youm black hole black hole background. We then specialized to circular supertubes and used the potential to search for bound states of supertubes and black holes.

For the near-extremal black hole, we found there are metastable and stable bound states (potential at the minimum higher or lower than its horizon value). Angular momentum of the black hole was found to be crucial for the existence of the repulsive forces which produce stable bound states. Either the black hole and the supertube rotate in the same plane and we observed stable bound states for large background angular momentum, oriented to give a repulsive interaction. Or the black hole and the supertube have rotation in orthogonal planes and then the repulsion is a result of the interaction of the supertube with the third charge and the angular momentum orthogonal to the supertube. Due to the large number of  background and supertube charges, we have concentrated on an example-based scan of this  space of parameters to get a qualitative view of when metastable, stable or no bound states appear. It would be interesting to perform a more detailed numerical analysis. This would allow a more quantitative view of the supertube--black hole phases and of how the phase boundaries depend on the mass, charges and angular momenta of the background. 

In section \ref{ss:Conclusions_Sect3}, we motivated that metastable bound states should always decay to a well-defined Cvetic-Youm black hole. In particular, it should be impossible to overspin the black hole by bringing in a supertube, such that naked closed timelike curves  (CTC's) do not form. The stable bound states on the other hand can either be the end product of the decay of a Cvetic-Youm black hole, or they can be new non-trivial solutions for a set of charges for which no CTC-free Cvetic-Youm black hole exist. This hints at an interesting black hole phase diagram. For a given set of charges and angular momenta, the black hole is not necessarily the unique solution -- there can also be supertube-black hole bound states.\footnote{Other two-center non-extremal solutions have been found in the literature where a black hole is immersed in scalar hair condensate and can be thought of as the end point of superradiance \cite{Gubser:2008px,Basu:2010uz,Bhattacharyya:2010yg}. The difference with our construction is that the the supertube--black hole bound state carries a dipole charge and cannot be the end point of superradiance.}  
This extends the black hole phase structure in five-dimensional general relativity (Myers-Perry black holes \cite{Myers:1986un}, black rings \cite{Emparan:2001wn,Pomeransky:2006bd}, helical black rings \cite{Emparan:2009vd} and combinations \cite{Elvang:2007rd,Iguchi:2007is,Evslin:2007fv,Izumi:2007qx,Elvang:2007hs}) to a theory with electric fields and Chern-Simons terms. 
It would be very interesting to study the black hole/bound state phases and decay processes in more detail, both in a probe and a back-reacted setup. 
Furthermore, if the back-reacted solutions persist all the way to the unitarity bound then they would correspond to adding non-extremality to the two-charge fuzzball solutions \cite{Lunin:2002qf}. It would then be interesting to study if small non-extremality can be added to the two-charge fuzzballs and the resulting solution coarse grained to produce the supertube--black hole bound states.

\medskip

We have also studied the same setup in the D1-D5 decoupling limit. This allowed an interpretation of the bound states in the dual D1-D5 orbifold CFT. We restricted to no net third charge (`P' in the D1-D5-P frame) and we have seen that a thermodynamical instability of the CFT (a new `enigmatic phase' being entropically favoured over the `black hole phase') maps qualitatively to the existence of a dynamical instability in the bulk (supertube--black hole bound states having lower energy than the configuration where the supertube falls to the black hole horizon). This is the non-extremal realization of the supersymmetric results of \cite{Bena:2011zw}. The existence region for stable bound states on the bulk side falls well within the region where the enigmatic phase is entropically dominant in CFT. The reason we find only a small part of the phase diagram in the supergravity analysis, is that we are bound by the validity of the probe approximation and the existence of the background black hole (cosmic censorship). We believe that for the back-reacted solution describing the bound state of the supertube and the non-extremal black hole, also the phase boundaries of the supergravity and the CFT regimes will map as they do in the supersymmetric setup of \cite{Bena:2011zw}. 

We noted that the metastable bound states that appear in the bulk have no direct interpretation in the orbifold CFT. It would be interesting to move away from the orbifold point and identify the dynamical process explaining the existence of metastable and stable new phases also in the CFT.

\medskip

Finally, our current treatment and the one of \cite{Bena:2011fc} show that the physics of supertubes is a very interesting tool to understand non-extremal solutions, much like it has been for supersymmetric solutions. As we mentioned before, their study could give more insight on the decay processes of non-extremal solutions. One could also study entropy enhancement of supertubes in non-extremal background (analogous to \cite{Bena:2008nh} for supersymmetric setups). There is still a lot of rich physics that can be obtained from further probe calculations. For instance, going back to the original motivation of our paper, it would be interesting to extend the current static analysis to scattering of supertubes off non-extremal black holes. We believe that one would observe a supertube analogue of the  Penrose  process. (See \cite{Marolf:2005cx} for a similar analysis in the supersymmetric BMPV background.)

\section*{Acknowledgments}

We thank I.\ Bena, F.\ Denef, S.\ Giusto, S.\ Mukhi, A.\ Virmani and N. Warner for stimulating discussions and the Pedro Parcual Centro de Ciencias in  Benasque and the organizers of the Benasque workshop on ``Gravity: perspectives from strings and higher dimensions'' for providing the stimulating environment where this work was initiated. BV thanks ITFA, University of Amsterdam, where part of this work was done, for hospitality. The work of BDC is supported by the ERC Advanced Grant 268088-EMERGRAV.  BV is supported by the ERC Starting Independent Researcher Grant 240210 - String-QCD-BH.

\appendix{
\addtocontents{toc}{\protect\setcounter{tocdepth}{1}}%

\section{Notations and conventions}
\label{app:Conventions}

For the largest part of this paper, we follow conventions of \cite{Bena:2008dw}, appendix D. 

Newton's constant in $D$ spacetime dimensions is related to the $D$-dimensional Planck length as
\be
G_D = (2 \pi)^{D-3} (\ell_D)^{D-2}\,.
\ee
The string length is $\ell_s^2 = \alpha'$. The tensions of the extended objects in string and M-theory are:
\eal{
T_{F1} = \frac 1 {2\pi \alpha'}\,,& \qquad T_{Dp} = \frac 1 {g_s (2\pi)^p (\ell_s)^{p+1}}\,,\qquad T_{NS5} = \frac 1 {g_s^2(2\pi)^5 (\ell_s)^6}\,,\\[1mm]
&T_{M2} = \frac 1 {(2\pi)^2 (\ell_{11})^3} \,,\qquad T_{M5} = \frac 1 {(2\pi)^5 (\ell_{11})^6}\,,
} 
where $g_s$ is the string coupling constant. It is related to the eleven-dimensional Planck length as
\be
\ell_{11} = g_s^{1/3} \ell_s\,.
\ee
In a compactification of M-theory along a circle of radius $R_{11}$, the IIA coupling constant is given as
\be
R_{11} = g_s \ell_s\,.
\ee
For convenience we list the masses of a single M2 brane wrapping the first, second or third torus in a $T^6$ compactification of M-theory:
\be
m_1=\f{R_5 R_6}{l_{11}^3}, \quad m_2=\f{R_7 R_8}{l_{11}^3}, \quad m_3=\f{R_9 R_{10}}{l_{11}^3}\,. \label{massCharges}
\ee
In a $T^6$ compactification of M-theory, where the radius of each torus circle is $R_5,\ldots ,R_{10}$, the five-dimensional Planck length is related to the eleven-dimensional Planck length as 
\be
G_5 = \frac{G_{11}}{vol(T^6)} = \frac{G_{11}}{(2\pi)^6 R_5 R_6 R_7 R_8 R_9 R_{10}} = \frac \pi 4 \frac{(\ell_{11})^9}{R_5 R_6 R_7 R_8 R_9 R_{10}} = \f{\pi}{4} \f{1}{m_1 m_2 m_3}\,. \label{G5}
\ee
The relation between the integer charges counting the number of M2 and M5 branes, $\bN_I$ and $n_I$, and the physical charges of the five-dimensional solution, $\bQ_I$ and $d_I$, upon compactification of M-theory on $T^6$ is 
\be
\bQ_1 =\f{\bN_1}{m_2 m_3} \,,\qquad \bQ_2 =  \f{\bN_2}{m_1 m_3}\,, \qquad \bQ_3 = \f{\bN_3}{m_1 m_2}\,,\qquad d_I = \frac{n_I}{m_I}\,.
\ee

\medskip

We adopt the conventions
\be
m_1=m_2=1, \, m_3= \epsilon \,.\label{CFTLimit1}
\ee
which leads to 
\be
\frac \pi {4 G_5} = \epsilon \,.\label{eq:5dNewton}
\ee
In the largest part of this paper, we put $\epsilon = 1$. The main exception is section \ref{s:Decoupling}, where we keep it explicit. Note that the equalities \eqref{CFTLimit1}, \eqref{eq:5dNewton} are numerical, but are not dimensionally correct. They should be interpreted as fixing the unit of length.

\section{Hamiltonian for supertubes in Cvetic-Youm background}
\label{app:Hamiltonian}

In this section we derive the Hamiltonian for an M2 -- M2 $\to$ M5 supertube in the eleven-dimensional non-extremal black hole background \eqref{eq:11d_Background}. The result applies to all duality frames which give the same five-dimensional geometry. As for the supertube in flat space \cite{Mateos:2001qs,Emparan:2001ux}, the Hamiltonian is derived from the Born-Infeld action of the dipole brane, in our case an M5 brane. Since the Born-Infeld description of an M5 brane is rather involved, we go to a duality frame more amenable to computations. We reduce along one of the torus coordinates such that the dipole brane is a D4 brane and the tube charges $\pQ_1$ and $\pQ_2$ become an F1 and a D2 respectively. From the D4 brane Born-Infeld action, we get the Hamiltonian by a Legendre transform. The calculation is similar to the one for supertubes in supersymmetric three-charge backgrounds \cite{Bena:2011fc}.

\subsection{Reduction to IIA over a torus direction}
We reduce over one of the directions of the first two-torus. The remaining coordinate of that $T^2$ is called $z$. The metric and dilaton are (in string frame)
\bea
ds_{10,str}^2 &=&  e^{2\Phi/3}\left[-Z^{-2} H_m (dt + k)^2 + Z ds_4^2 + \frac{Z}{ H_1} dz^2 + \frac{Z}{ H_2}ds_{T_2}^2+ \frac{Z}{ H_3}ds_{T_3}^2\right]\,,\label{eq:10d_metric}\nn
e^{2\Phi} &=& (Z/H_1)^{3/2}=  \sqrt{\frac{H_2 H_3}{H_1^2}}\,.
\eea
The gauge fields are
\bea
B_2 &=& A^{(1)} \wedge dz\,,\nn
C_3 &=& A^{(2)} \wedge dT_2 + A^{(3)} \wedge dT_3\,. \label{C3}
\eea
We also need the following components of the five-form gauge field:
\bea
C^{(5)}_{t \psi y 34} &=&  \f{m \cos^2 \theta}{f H_1} (a_2 c_1 c_2 s_3 - a_1 s_1 s_2 c_3)\,, \\
C^{(5)}_{t \phi y 34} &=&  \f{m \sin^2 \theta}{f H_1} (a_1 c_1 c_2 s_3 - a_2 s_1 s_2 c_3)\,.
\eea
The five-form potential can be obtained from $C_3$ by Hodge dualization as $d C_5 = -\star d C_3 + H_3 \wedge F_3 $, or by the appropriate duality chain from the RR gauge field $C_2$ in the original Cvetic-Youm solution, which is in a frame where the three charges correspond to D1 branes, D5 branes and momentum (P) along their common direction. For the form of $C_2$ in that frame, see \cite{Giusto:2004id}.

\subsection{DBI Lagrangian}

We consider the embedding of a D4 brane, with dissolved D2 and F1 charges, in the ten-dimensional background. We wrap the D4 brane on the $z$-direction, on the second $T^2$ and along a one-cycle in the external space. With embedding coordinates $\xi^0\ldots \xi^4$, we have
\be
t = \xi^0\,,\quad z = \xi^1\,,\quad \psi = b_1 \xi^2\,,\quad \phi = b_2 \xi^2\,, \label{worldVolumeCoordinates}
\ee
and $\xi^3,\xi^4$ make up the second two-torus. To avoid unnecesary confusion in subscripts, we single out the $\xi^2$ coordinate as
\be
\a \equiv \xi^2\,.
\ee

The world-volume field on the D4 brane is:
\be
2 \pi \alpha'F = \cale d\xi^0 \wedge d \xi^1 + \calb d\xi^1 \wedge d\a\,.
\ee
The magnetic field $\calb$ is a source for D2 charge inside the D4 worldvolume. The electric field $\cale$ is a source for F1 charge.

\medskip 

The action for a D4 brane wrapped $|N_{D4}|$ times in the black hole background is
\bea
S &=& S_{BI} +S_{WZ}\,,\nn
S_{BI} &=& - |N_{D4}| T_{D4}\int d^5 \xi\ e^{-\Phi}\sqrt{-\det{(g + B + 2 \pi \a' F)}}\,,\nn
S_{WZ} &=&  N_{D4} T_{D4}\int C_5 - N_{D4} T_{D4}\int(B + 2 \pi \a' F)\wedge C_3\,,
\eea
where all fields are interpreted as pull-backs on the D4 worldvolume. If the brane is an anti-D4 brane then $N_{D4}$ is negative. 

After some algebra, the Born-Infeld action is
\bea
S_{BI}  &=& - m_1 m_2 \,|N_{D4}| \int d \xi^0 \frac{Z}{H_2}\sqrt{ g^{(4)}_{\a\a} \frac{Z H_m }{H^2_1} + \frac{H_m}{ Z^2}( \tilde \calb + k_\a \tilde \cale)^2 - g_{\a\a}^{(4)} Z \tilde \cale^2}\,,
\eea
where the proportionality constants $m_1,m_2$ are the masses of the first and second M2-brane (see eq.\ \eqref{massCharges}) and satisfy:
\be
m_1 m_2 = T_{D4}\int d \xi^1 d\a d\xi^3 d \xi^4 =T_{D4} R_z (2\pi)(2\pi R_5)(2\pi)(2\pi R_6)\,,
\ee
and we defined the shifted electric and magnetic fields
\bea
\tilde \cale &=& \cale + B_{01} = \cale + A^{(1)}_0 = \cale + \coth \delta_1 \,(H_1^{-1}-1)\,,\\
\tilde \calb &=& \calb + B_{12} = \calb - A^{(1)}_\a = \calb - \coth \delta_1 \,H_1^{-1} k_\a - B_\a^1\,.
\eea
When the background has no rotation ($k_\a=0$), this is exactly the structure of the supertube in flat space. 

The Wess-Zumino action is 
\bea
S_{WZ} &= m_1 m_2 N_{D4} \int d \xi^0\Big{[} &\tilde \cale  A^{(2)}_\alpha  + \tilde \calb A^{(2)}_0
+\coth \delta_1 \coth \delta_2  \frac{k_\alpha}{H_1 H_2}  - \frac{\coth \d_1}{H_1}\ A^{(2)}_\alpha \nn
&&- \coth \d_2\ A^{(1)}_\alpha
+\frac{c_3}{s_1 s_2} \frac{m}{f H_m} (a_1 b_1 \cos^2 \theta + a_2 b_1 \sin^2 \theta)\Big{]}\,.
\eea

\subsection{Hamiltonian}
The D4 brane action describing the F1--D2 $\to$ D4 supertube contains the conserved D2 charge, $\calb$, but the electric field is not a conserved quantity. To find the Hamiltonian, which depends on the conserved D2 and F1 charges, we need to perform the Legendre transform of the action with respect to $\cale$. 

The Hamiltonian density is defined as
\be
\calh = \cale \frac{\pa \call }{\pa \cale } - \call\,.
\ee
The physical charges $\pQ_1,\pQ_2$ (dimensions of length squared) and dipole charge $d_3$ (dimension of length) of the supertube are (we use \eqref{massCharges}):
\be
\pQ_1 \equiv \f{4 G_5}{\pi} \,  \f{\partial \call}{\partial \cale} \,,\qquad \pQ_2 \equiv d_3 \calb\,,\qquad d_3 \equiv \frac{N_{D4}}{m_3}\,.
\ee

Then the Hamiltonian is
\bea
\f{4 G_5}{\pi} \calh &=& \f{1}{|d_3|} \frac{\sqrt{H_m Z^3 g^{(4)}_{\a \a}}}{Z^3g^{(4)}_{\a \a} - H_mk_\a^2}\sqrt{\left(\ti \pQ_1 \right)^2 + d_3^2 \frac{Z^3g^{(4)}_{\a \a} - H_m k_\a^2}{H_2^2}}\sqrt{\left(\ti \pQ_2\right)^2+ d_3^2  \frac{Z^3g^{(4)}_{\a \a} - H_m k_\a^2}{H_1^2}}\nonumber\\
&&+\f{1}{d_3}\frac{H_m k_\a}{Z^3g^{(4)}_{\a \a} - H_m k_\a^2} \ti \pQ_1 \ti \pQ_2  -\coth \d_1 \frac{ \ti \pQ_1  }{H_1} -\coth \d_2 \frac{ \ti \pQ_2}{H_2}  -\f{1}{d_3} \coth \d_1 \coth \d_2 \frac{k_\a}{H_1 H_2}\nn
&& - \f{1}{d_3}\frac{c_3}{s_1 s_2} \frac{m}{f H_m} (a_1 b_1 \cos^2 \theta + a_2 b_2 \sin^2 \theta ) + \coth \delta_1\, \pQ_1 + \coth \delta_2 \, \pQ_2 \,,
\label{eq:NonExtremalHam}
\eea
with the notation for the shifted charges
\bea
\ti \pQ_1 &\equiv& \pQ_1 -  d_3 A_\a^{(2)} = \pQ_1 - d_3 (B_\a^{(2)} +  ch_2 \frac{k_\a}{H_2})\,, \nn
 \ti \pQ_2 &\equiv& \pQ_2 - d_3 A_\a^{(1)} = \pQ_2 -  d_3 (B_\a^{(1)} + ch_1 \frac{k_\a}{H_1})\,. \label{ShiftedCharges_2}
\eea

\subsection{Symmetries of the Hamiltonian}\label{app:Scaling}

Because of symmetries in the equations of motion (both in supergravity and in the probe approximation), the Hamiltonian \eqref{eq:NonExtremalHam} also has several symmetries. We list them here.

\subsubsection{Continuous symmetries} 

Under a conformal rescaling of all length scales, the supergravity charges are affected, since the electric charges and angular momenta are dimensionful. The Hamiltonian scales as:
\be
\calh(m,\bQ_I,J_i;\pQ_J,d_3;r) = \frac 1 \lambda \calh(\lambda ^2 m, \lambda^2 \bQ_I,\lambda^3 J_i;\lambda^2 \pQ_J,\lambda d_3;\lambda r)\,,\label{eq:Scaling_1}
\ee
where $J_i$ denote the angular momenta of the black hole.

We can also consider the effect of a linear scaling of all integer charges, keeping length scales fixed. Since the probe Hamiltonian is first order in the probe charges, a linear scaling of the probe charges has the effect:
\be
\calh(m,\bQ_I,J_i;\pQ_J,d_3;r) = \frac 1 \lambda \calh(m, \bQ_I,J_i;\lambda\pQ_J,\lambda d_3;r)\,.
\ee
Note that there is no similar linear scaling of the background charges. This is because such a linear scaling would scale up all M-theory charges and there is a (hidden) Kaluza-Klein monopole charge in the four-dimensional base metric \eqref{eq:4d_Base}, which has been fixed to a certain value and is not a free parameter.\footnote{Note that the scaling \eqref{eq:Scaling_1} does {\em not} affect the kaluza-Klein monopole charge.}

In the D1-D5 frame  decoupling limit, $m_1 m_2 \gg m_3$, we can also consider the scaling of  various charges with the string coupling constant. This gives for the shifted Hamiltonian ${\mathscr H} = \calh - (\pQ_1 +\pQ_2)$:
\be
{\mathscr H}(m,\bQ_1,\bQ_2,\bQ_3,J_i;\pQ_J,d_3;r) = \frac 1 {g^2} \calh(g ^2 m, g \bQ_1, g\bQ_2, g^2\bQ_3,g^2 J_i;g \pQ_J, d_3;g \,r)\,,\label{eq:Scaling_2}
\ee
This scaling is not visible outside the decoupling limit because in general $m$ gets contributions from all three charges and has no clean scaling. In the decoupling limit, $m$ only has contributions related to the third charge and $m$ scales in the above way.

\subsubsection{Discrete symmetries}

First we consider only the Cvetic-Youm background \eqref{eq:11d_Background}. We can perform several $\mathbb{Z}_2$ transformations that give an equivalent physical background. For example, flipping the sign of  $\bQ_I$ is equivalent to $(\delta_I,a_1,\phi) \to (-\delta_I,-a_1,-\phi)$. This means one obtaines an equivalent background with reversed charge of type ``$I$'' by performing $\bQ_I\to -\bQ_I$ and $J_\psi \to -J_\psi$. 

When we consider probe supertubes, only some of these sign choices give inequivalent results. In this paper, we use two discrete symmetries of the Hamiltonian. One that flips the signs of $\bQ_1,\bQ_2$ of the background and $\pQ_1,\pQ_2$ of the probe, and on that affects the pair $(\bQ_3,J_\phi$ of the background:
\bea
\calh(m,\bQ_1,\bQ_2,\bQ_3,J_\phi,J_\psi;\pQ_1,\pQ_2,d_3;r) &=&\calh(m, -\bQ_1,-\bQ_2,\bQ_3,J_\psi,J_\phi;-\pQ_1,-\pQ_2,d_3;r)\nn
&=&\calh(m, \bQ_1,\bQ_2,-\bQ_3,J_\psi,-J_\phi;\pQ_1,\pQ_2,d_3;r)\,.
\eea

\section{D1-D5 CFT} \label{section:D1D5CFT}
In this section we give a quick review of the D1-D5 CFT\@. For a more
detailed review, see for example \cite{David:2002wn}.

Consider type IIB string theory on $S^1 \times M^4$ with $\bN_1$ D1-branes
wrapping $S^1$ and $\bN_5$ D5-branes wrapping $S^1 \times M^4$, where
$M^4$ is $T^4$ or $K3$.  We take the size of $M^4$ to be string scale. The
Higgs branch of this system flows in the IR to an $\mathcal N=(4,4)$
SCFT whose target space is a resolution of the symmetric product
orbifold $\mathcal M=(M^4)^{\bN}/S_{\bN}\equiv {\rm Sym}^{\bN}(M^4)$, where $S_{\bN}$ is
the permutation group of order $\bN$ and $\bN=\bN_1 \bN_5$ ($\bN=\bN_1 \bN_5+1$) for
$M^4=T^4$ (for $M^4=K3$). The orbifold $\mathcal M$ is called the
``orbifold point'' in the space of CFT's and the theory is easy to analyze at that point.

This CFT is dual to type IIB string theory on $AdS_3 \times S^3 \times
M^4$. To have a large weakly-coupled $AdS_3$, $\bN$ must be large and the
CFT must be deformed far from the orbifold point by certain marginal
deformations (for recent work see \cite{Avery:2010er,Avery:2010hs,Avery:2010vk}). 

For presentation purposes, we will henceforth take $M^4=T^4$, but much
of the discussion goes through also for $M^4=K3$. 
The theory has an $SU(2)_L \times SU(2)_R$ $R$-symmetry which originates
from the $SO(4)$ rotational symmetry transverse to the D1-D5 system.
There is another $SU(2)_1 \times SU(2)_2$ global symmetry which is
broken by the toroidal compactification but can be used to classify
states. We label the charges under these symmetries as $\alpha,\dot
\alpha$ and $A,\dot A$ respectively. At the orbifold point each copy of
the CFT has four left-moving fermions $\psi^{\alpha A}$, four left-moving bosons $\partial X^{A \dot A}$, four right-moving fermions
$\psi^{\dot \alpha A}$ and four right-moving bosons $\bar \partial X^{A
\dot A}$. In addition the CFT has twist fields $\sigma_n$ which
cyclically permute $n \le \bN$ copies of the CFT on a single $T^4$. One
can think of these twist fields as creating winding sectors in the D1-D5
worldsheet with winding over different copies of the $T^4$.

The D1-D5 CFT is in the  Ramond-Ramond sector because of asymptotic flatness and supersymmetry. Elementary bosonic twist fields (without any
bosonic or fermionic excitations) are charged under $SU(2)_L \times
SU(2)_R$ {\it viz.}\ $\sigma_n^{\alpha \dot \alpha}$ or under $SU(2)_1 \times
SU(2)_1$ {\it viz.}\ $\sigma_n^{A B}$ while elementary fermionic twist
fields are charged under $SU(2)_L \times SU(2)_1$ {\it viz.}\ $\sigma_n^{\alpha
A}$ or $SU(2)_R \times SU(2)_1$ {\it viz.}\ $\sigma_n^{\dot \alpha A}$. A
general Ramond sector ground state is made up of these bosonic and
fermionic twist fields with the total twist $\sum n=\bN$ as
\begin{gather}
 |gr,gr \rangle = \prod_{n,\alpha,\dot \alpha, A,\dot A} (\sigma_n^{\alpha \dot \alpha})^{\bN_{n,\alpha\dot \alpha}} (\sigma_n^{A B})^{\bN_{n, A B}} (\sigma_n^{\alpha A})^{\bN_{n ,\alpha A}} (\sigma_n^{\dot \alpha A})^{\bN_{n,\dot \alpha A}}, \nonumber \\
 \sum_{n,\alpha,\dot \alpha, A,B}  n( \bN_{n,\alpha\dot \alpha} + \bN_{n,AB} + \bN_{n,\alpha A}+\bN_{n,\dot \alpha A})=\bN, \nonumber \\
\qquad \bN_{n,\alpha\dot \alpha}=\bN_{n,AB}=0,1,2,\dots, \quad \bN_{n,\alpha A}=\bN_{n,\dot \alpha A}=0,1.
\end{gather}

A general Ramond sector state is made of left- and right-moving excitations on the Ramond ground states
\begin{equation}
 |ex,gr\rangle, \qquad |gr,ex\rangle, \qquad |ex,ex \rangle
\end{equation}
where ``$ex$'' means acting on Ramond ground states ``$gr$'' by the bosonic and
fermionic modes. In Fig.~\ref{fig:D1-D5states} we diagrammatically
represent a  non-SUSY state in the CFT with both left- and right-movers.  The arrows represent different $R$-charges of
elementary twists.
\begin{figure}
\begin{center}
   \includegraphics[width=7cm]{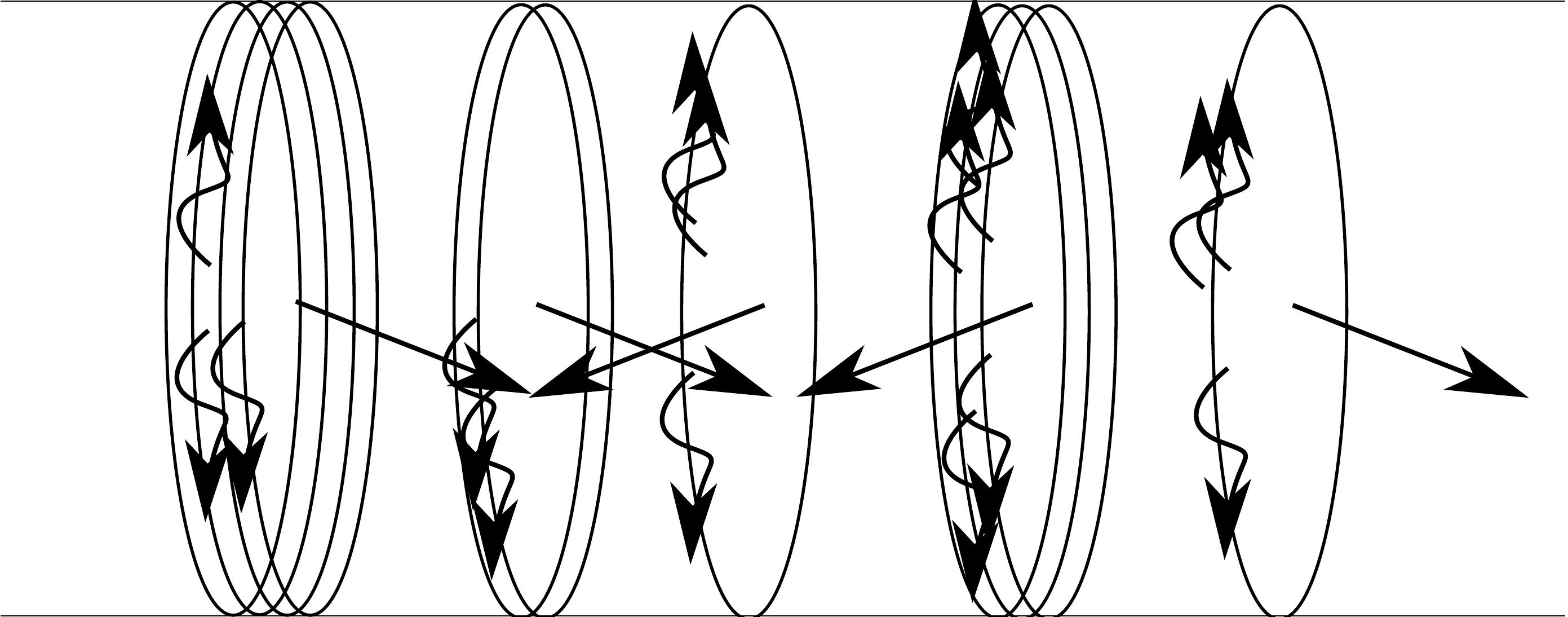}
 \caption{\sl Various states in the Ramond sector of the D1-D5 CFT.\label{fig:D1-D5states}}
\end{center}
\end{figure}
The states of the CFT are characterized by their left and right
dimension ($L_0$ and $\bar L_0$) and $R$-charges ($J_L$ and $J_R$). 
In our conventions, $J_{L,R}$, the third components of the $SU(2)_{L,R}$ generators $\vec
J_{L,R}$ are half-integers.  The Ramond
sector ground states all have the same dimension $L_0 = \bar
L_0=\f{\bN}{4}$. An excited state has dimension greater than that of the
ground state and any additional dimension is related to the left- and
right-moving momentum along the branes by
\be
\bN_{pL}= L_0 - \f{\bN}{4}, \qquad  \bN_{pR} = \bar L_0 - \f{\bN}{4} 
\ee
The relation of the CFT quantities with the bulk quantities are
\begin{gather}
  \f{ m \cosh 2 \delta_p}{2} =\f{l_{11}^6}{V_{T1} V_{T2}} (\bN_{pL} + \bN_{pR}) , \qquad \f{m \sinh 2 \delta_p}{2} =\f{l_{11}^6}{V_{T1} V_{T2}}  (\bN_{pL} - \bN_{pR}) \nn
\f{\pi}{4 G_5} J_\phi= J_R-J_L, \qquad \f{\pi}{4 G_5} J_\psi= -J_R - J_L \,,\label{AdSCFTmap}
\end{gather}
with the identification $\bN_{pR} =\bN_{3R},\,\bN_{pL}=\bN_{3L}, \delta_p=\delta_3$.

\section{Relation to ``Hot Halos and Galactic Glasses''}
\label{app:Comparison_Denef}

We explain the connection to \cite{Anninos:2011vn}. Those authors consider four-dimensional black holes of IIA string theory. They put D6-D4-D2-D0 probes in the non-extremal, static, D4-D0 black hole background.  In this appendix, we show how their background and probes can be  connected to ours, and what the similarities and differences of the resulting physics are.

First, we explain the dualization of the non-extremal D0-D4 background to the D6-D2 frame, which we then uplift to eleven-dimensions. This gives the non-rotating Cvetic-Youm black hole of section \ref{s:Background+Hamiltonian}. Afterwards, we compare the probes. We conclude with some remarks on the (dis)similarities.

\subsection{Background}

\subsubsection*{Four-dimensional non-extremal D0-D4 black hole}

In \cite{Anninos:2011vn}, the static non-extremal D0-D4 black hole background \cite{Gibbons:1982ih,Galli:2011fq} is considered. For later comparison, we write down their solution as a truncation of a  $T^6$ compactification of IIA string theory, known as the STU model. This gives $N=2$ supergravity coupled to three vector multiplets: there are 3 scalar fields $z^A$ and 4 vector fields  $\hat A_\mu^0,\hat A_\mu^A$ (including the graviphoton). In \cite{Anninos:2011vn} a further truncation is made to one vector multiplet as\footnote{We will use hats on coordinates, charges and functions whenever confusion with those of the five-dimensional Cvetic-Youm solution is possible.}
\be
z^A = K^A \hat z^1\,,\qquad \hat A^A_{\mu} = K^A \hat A^1_{\mu}
\ee
where $K^A$ are constants normalized to
\be
D \equiv 6 K^1 K^2 K^3\,.
\ee
In \cite{Anninos:2011vn}, $D$ is put to one. We keep it explicit for now and will put $D=2$ later.

The four-dimensional black hole metric geometry is of the form
\be
ds_4^2 = -e^{2U} dt^2 + e^{-2U} \left(\frac{c^4}{\sinh ^4 c \tau} d \tau^2 + \frac{c^2}{\sinh ^2 c \tau} (d\hat \theta^2 + \sin^2 \hat \theta d \hat\phi^2)\right)\,,
\ee
with spatial infinity at $\tau = 0$, and the scalar field $\hat z^1$ is purely imaginary:
\be
\hat z^1 = i y\,.
\ee

The solution is given in terms of the functions
\be
\hat H_0 =\frac{|\hat Q_0|}{c} \sinh (c \tau + c_2)\,,\qquad
\hat H_1 =\frac{|\hat P_1|}{c} \sinh (c \tau + c_4)\,.
\ee 
The equations of motion determine $U$ and $y$ as
\bea
e^{-2 U} &=& \sqrt{\frac{2}{3} \hat H_0 \hat H^3_1}\,,\qquad y = \sqrt{\frac{6 \hat H_0}{\hat H_1}} \,,
\eea
and the gauge potentials are
\be
\begin{array}{lcl}
\hat A^0= \frac{1}{2 \hat Q_0}\left(\sqrt{c^2 + \frac{\hat Q_0^2}{\hat H_0^2}} - c\right) dt\,,&\qquad& \hat A^1 = \hat P_1 (1-\cos\hat \theta)d \hat \phi\,,\\
\hat B_0=\hat Q_0(1-\cos \hat\theta)d\hat\phi\,,&\qquad & \hat B_1 = -\frac{3}{2\hat P_1}\left(\sqrt{c^2 + \frac{\hat P_1^2}{\hat H_1^2}}-c\right)dt\,.
\end{array}
\ee
We have included the dual gauge potentials $\hat B_0,\hat B_1$, for more information see \cite{Anninos:2011vn}. 

Note that the constants $c_2$ and $c_4$, which determine the asymptotic values of the functions $\hat H_0,\hat H_1$, should be fixed by asymptotic boundary conditions. In \cite{Anninos:2011vn}, four-dimensional asymptotics are chosen. We leave the constant $c_2$, which sets the string coupling constant, unfixed, to interpolate between weakly coupled IIA string theory (four non-compact dimension) and M-theory (five non-compact dimensions).

\subsubsection*{Uplift to M-theory}

We perform the dualization to the eleven-dimensional M2-M2-M2 black hole in three steps. First we uplift the four-dimensional solution to IIA. Then we dualize it to a D2-D6 black hole. The D6 charge then makes an uplift to eleven dimensions possible.

The uplift of the D0-D4 solution to the IIA string frame metric is
\be
\widetilde{ds}_{10}^2 = ds^2_4 + \sum_{A=1}^3 y K^A ds_A^2 \,, \qquad e^{2\tilde \Phi} =   K^1 K^2 K^3 y^3 = \frac D 6 y^3\,,
\ee
where $ds_A^2$ are unit metrics on the two-tori $T_A$. We choose $C_5,C_7$ as independent RR fields:
\bea
C_5 = D^{ABC}(K^2)_A\hat B_1 \wedge \omega_B \wedge \omega_C\,,&\qquad& C_7 = D \hat B_0\wedge \omega_1 \wedge \omega_2 \wedge \omega_3\,.
\eea
where $(K^2)_A \equiv D_{ABC} K^B K^C$ and $D_{ABC} = D^{ABC}=|\epsilon_{ABC}|$ and $\omega_A$ are unit volume forms on the three orthogonal $T^2$'s inside $T^6$.

We perform 6 T-dualities along $T^6$. This gives a D6-D2 black hole in what we call the IIA$'$ frame. The ten-dimensional string frame solution is
\bea
ds_{10}^2 &=& ds^2_4 +  \sum_{A=1}^3 \frac{ ds_A^2} {y K^A} \,,\qquad e^{2 \Phi} = \frac 6 D y^{-3}\,,\nn
C_1 &=& D \hat B_0\,, \qquad C_3 = \hat B_1 \wedge (K^2)_A \omega_A \,.
\eea

The uplift to the eleven-dimensional metric is given as:
\be
ds_{11}^2 = e^{4\Phi /3} (d \varphi_{11} + C_1)^2  + e^{-2 \Phi/3} ds^2_{10}\,.
\ee
We make the redefinition $m = 4\ell_{11} \,c$ and the coordinate transformation
\be
\frac{m}{r^2} = 2 e ^{-c\tau}\sinh c\tau = 1 - e^{-2 c \tau}\,,
\ee
and we put $D = 2$ in the following. Then we define the functions
\be
H_A = K^A H_m^{1/2}\hat H_1\,,\qquad H_0 = 8 H_m^{1/2}\hat H_0\,,\qquad H_m = 1 - \frac m{r^2}\,.
\ee
All of these functions have the form $a + \frac b {r^2}$. We then find that the eleven-dimensional metric and three-form potential are
\bea
ds_{11}^2 &=& -(H_1 H_2 H_3)^{-2/3} H_m dt^2 + (H_1 H_2 H_3)^{-2/3} ds_4^2 + \sum_{A=1}^3 \frac{(H_1 H_2 H_3)^{-2/3}}{H_A} ds_A^2\,,\nn
A_3 &=& -\sum_A\frac{e^{c_4}}{\hat H_A}dt \wedge \omega_A\label{eq:11dMetric_Denef}
\eea
with four-dimensional base metric
\be
ds_4^2 = \frac 1 {H_0} (d \hat \psi + 2 D \hat Q_0 (1-\cos \hat \theta) d\hat \phi)^2 + H_0 (\frac{r^2 dr^2}4 + \frac{r^4}{16} (d \hat \theta^2 + \sin^2 \hat \theta d \hat \phi^2))\,,
\ee
with $\hat \psi = 2 \varphi_{11}$ has period $4\pi$.

\begin{table}[ht!]
\centering
\begin{tabular}{|l||c|c|c||c|c|}
\hline
&& IIA&IIA$'$
& \multicolumn{2}{c|}{M}\\
\hline
\hline
Background & $\hat P^A$ &  D4 ($T_B \times T_C$) & D2($T_A$) & $\bQ_A$ & M2($T_A$)\\
 & $\hat Q_0$ &  D0  & D6($T^6$)   &  $\bQ_0$ (fixed) & KKmon\\
\hline
\hline
Probes    &$\hat p^0$ & D6&D0& $J_\a$&Ang.\ Mom.\\\
&$\hat p^A$&D4($T_B \times T_C$) &D2 ($T_A$)&$\pQ_A$&M2 ($T_A$)\\
&$\hat q_A$&D2 ($T_A$)&D4($T_B \times T_C$) &$d_B$&M5($T_B\times T_C \times S^1_\a$)\\
&$\hat q_0$&D0&D6 ($T^6$)&/&/\\
\hline
\end{tabular}
\caption{Overview of the background and probe charges in the IIA/four-dimensional language of \cite{Anninos:2011vn} and in the M-theory/five-dimensional language used in this paper. We compare the charges and notations and on which two-tori $T_A$ they are wrapped. The coordinate $\a$ is along the M-theory circle, $P^A \equiv K^A \hat P_1$. Note that in the probe limit the charge in the bottom line cannot be mapped  to a probe charge in five dimensions.}\label{tab:Charges_Comparison}
\end{table}

\subsubsection*{4D/5D connection}

The eleven-dimensional solution \eqref{eq:11dMetric_Denef} is very similar to the non-rotating Cvetic-Youm solution \eqref{eq:11d_Background}. However, at this point it has only four non-compact directions. The asympotic form of the metric \eqref{eq:11dMetric_Denef} is set by the value of $H_0$ at spatial infinity.  When $H_0 = cst$ at spatial infinity ($c_2 \neq0$), the four-dimensional base is asymptotic to $\mathbb{R}^3 \times S^1$: spacetime is compactified to the four-dimensional black hole of \cite{Anninos:2011vn}. When $H_0 = 0$ at spatial infinity ($c_2=0$) there is a fifth non-compact direction. By choosing the value of the charge $\hat Q_0$  and the constant $c_4$ as\footnote{In conventional coordinates $\rho = r^2/4$, we have $H_0 = 4 \hat Q_0/\rho$ and $\hat Q_0=\ell_{11}/4$ gives one unit of Kaluza-Klein monopole charge.}
\bea
\hat Q_0 = \ell_{11}/4\,,\qquad e^{c_4} = \coth \delta_1\,,
\eea
the four-dimensional base space becomes $\mathbb{R}^4$ and the eleven-dimensional solution is the Cvetic-Youm black hole \eqref{eq:11d_Background} with  $J_\psi = J_\phi = 0$ and electric charges $\bQ_A = 2 K^A {\hat P_1}$.  This is the same 4D/5D connection as for supersymmetric multi-center black hole and black ring solutions \cite{Gaiotto:2005gf,Gaiotto:2005xt,Elvang:2005sa,Bena:2005ni,Behrndt:2005he,Ford:2007th}. For an overview of the relation between the charges in four and five dimensions, see Table \ref{tab:Charges_Comparison}.

\subsection{Probe}

In the D0-D4 frame, the authors of \cite{Anninos:2011vn} consider probes with D6-D4-D2-D0 charges. In the D6-D2 frame, these are again of the same form. They cannot be uplifted to eleven dimensions, as the D6 probe brane charge does not have a higher-dimensional probe analogue. This is because D6 charge maps to a geometric charge (Kaluza-Klein monopole), which cannot be treated  as a probe, in contrast to electrically or magnetically charged probe branes. For an overview of the probes in the different frames, see Table \ref{tab:Charges_Comparison}.

The problematic D6 charge (in the D6-D2 frame) is a D0 charge in the D0-D4 frame used in \cite{Anninos:2011vn}. We see that one needs to truncate this charge ($q_0 = 0$ in \cite{Anninos:2011vn}) to be able to compare to brane probes in the five-dimensional non-extremal black hole background.


\subsection{Comments}

In conclusion, we see that the uplift of the D0-D4 background of \cite{Anninos:2011vn} to five non-compact dimensions gives the Cvetic-Youm black hole without angular momentum ($J_\psi = J_\phi =0$ or $a_1=a_2=0$). Moreover, the D6-D4-D2-D0 probes of \cite{Anninos:2011vn} can only be uplifted if we truncate one of the charges (D0 probe charge in the D0-D4 background frame).

Therefore, both treatments are complementary. First, the geometries have different intrinsic dimensionalities (four-dimensional vs.\ five-dimensional). Second, both treatments have features the other one does not have. The rotating Cvetic-Youm background has two angular momenta, which would map to extra D6 charge and four-dimensional angular momentum in the D0-D4 background. And the probes in the four-dimensional background have an additional charge (the charge that cannot be uplifted in the probe limit) compared to the supertube probes we use in this paper. Moreover, supertube probes are smooth upon backreaction, while the brane probes of \cite{Anninos:2011vn} generically describe charges that would backreact to black holes. Also the origin of the attractive interaction responsible for stable bound states is complementary as these are largely related to exactly those charges that do not map to the other frame (five-dimensional angular momentum in our frame, D0-charge and compactification in the D0-D4 frame of \cite{Anninos:2011vn}).

}


\providecommand{\href}[2]{#2}\begingroup\raggedright\endgroup

\end{document}